\documentclass[a4,showpacs,twocolumn, amsmath, aps, floatfix, superscriptaddress]{revtex4-1}
\usepackage{hyperref}
\usepackage{graphicx,color}
\usepackage{amssymb}
\usepackage{dcolumn}
\usepackage[normalem]{ulem}
\newcommand{\tfav}{\langle F_z(z)\rangle }
\hypersetup{colorlinks=true,citecolor=blue}

\begin{document}
\title{Fast cholesterol flip-flop  and lack of
swelling in skin lipid multilayers}
 \author{Chinmay Das}
 \email{c.das@leeds.ac.uk}
 \affiliation{School of Physics and Astronomy, University of Leeds,
 Leeds LS2 9JT, United Kingdom}
 \author{Massimo G. Noro}
\email{massimo.noro@unilever.com}
\affiliation{Unilever R\&D Port Sunlight, Quarry Road East, Bebington, Wirral, CH63 3JW, UK}
 \author{Peter D. Olmsted}
 \email{pdo7@georgetown.edu}
 \altaffiliation[Permanent Address: ]{Department of Physics, 37th and O Streets NW, Georgetown University, Washington DC 20057, USA}
 \affiliation{School of Physics and Astronomy, University of Leeds,
 Leeds LS2 9JT, United Kingdom}
\date{\today}

\begin{abstract}
Atomistic simulations were performed on hydrated model lipid multilayers that are representative of the  lipid matrix in the outer skin (stratum corneum). 
We find that cholesterol transfers easily between
adjacent leaflets belonging to the same bilayer via fast orientational diffusion
(tumbling) in the inter-leaflet disordered region, while at the same time there is a large free energy cost against swelling. This fast flip-flop may play
an important role in accommodating the variety of curvatures that would be 
required in the three dimensional arrangement of the lipid multilayers in skin, 
and for enabling  mechanical or hydration induced strains without large
curvature elastic costs.
\end{abstract}
\maketitle
\noindent{\bf Introduction:} The outer layer of skin
\cite{freinkel.skin.01}, called the stratum corneum (SC), comprises
non-viable corneocyte cells within a matrix of lipid multilayers, and
is the main barrier against water loss and uptake of foreign pathogens
and chemicals \cite{elias.sc.rev.05}. In a simplified {\em brick and
  mortar} picture \cite{michaels.sc.brick.75} the corneocytes are the
`bricks' and the lipid multilayers constitute the `mortar'. Possibly
because of the extensive work on phospholipid biomembranes, SC lipid
multilayers have often been considered to behave similarly to
phospholipids, with highly hydrophilic head groups that lead to a
bulk-like water layer between adjacent bilayers \cite{aberg.08,
  *moghimi.jpharm.96, *sparr.col.00}.
However, direct experimental support for  hydrated multilayers has been scant 
\cite{bouwstra.jid.91,*mak.pharmres.91,*kiselev.EBJ.05}. 

To explain  {\em in~vivo} and {\em in~vitro} structural data, a number of
detailed scenarios have been proposed  
 in which the lipids in multilayers are in crystalline or gel states, with
negligible diffusion \cite{swartzendruber.jid.89,
forslind.adv.94, bouwstra.jlr.98, mcintosh.bpj.03, hill.bba.03,
schroter.bpj.09, iwai.jid.12}. Yet, SC lipid multilayers  \textit{in vivo}
necessarily undergo large deformation during hydration/dehydration and 
mechanical deformation. For example,  the  
corneocyte  volume (diameter $\sim30\mu$m and thickness $\sim 300$~nm)
  can change  under hydration by up to a factor of three 
\cite{bouwstra.jid.03,richter.appphysa.01}. Moreover, the concomitant changes in local 
curvature of an adjacent gel-like lipid multilayer would require large elastic (or plastic) stresses.

\begin{figure}[bthp]
\centerline{\includegraphics[width=\linewidth,clip=]{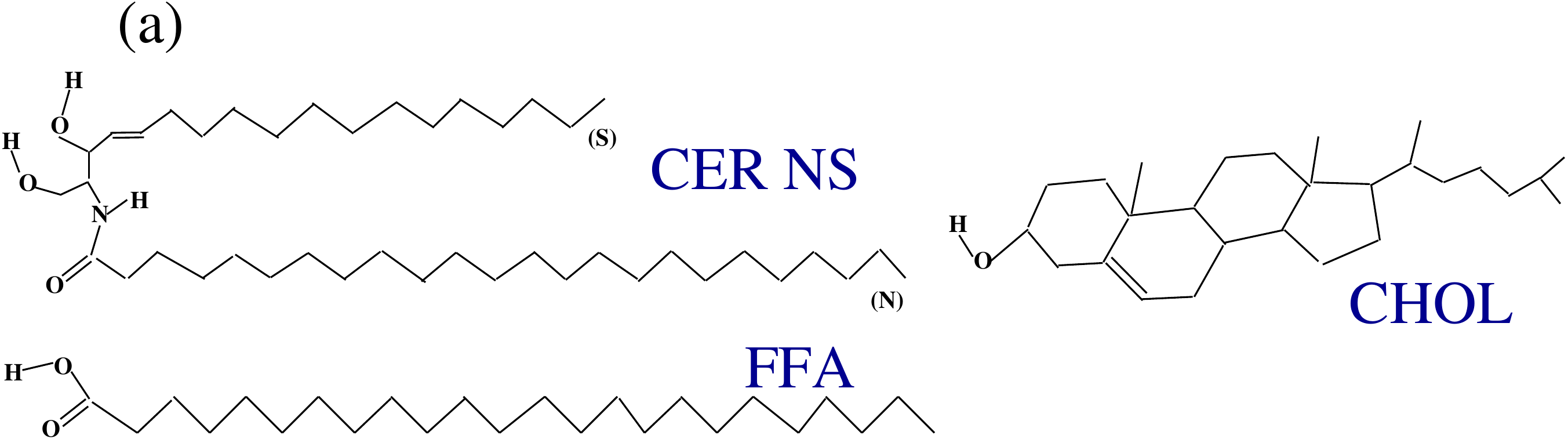}}
\centerline{\includegraphics[width=\linewidth,clip=]{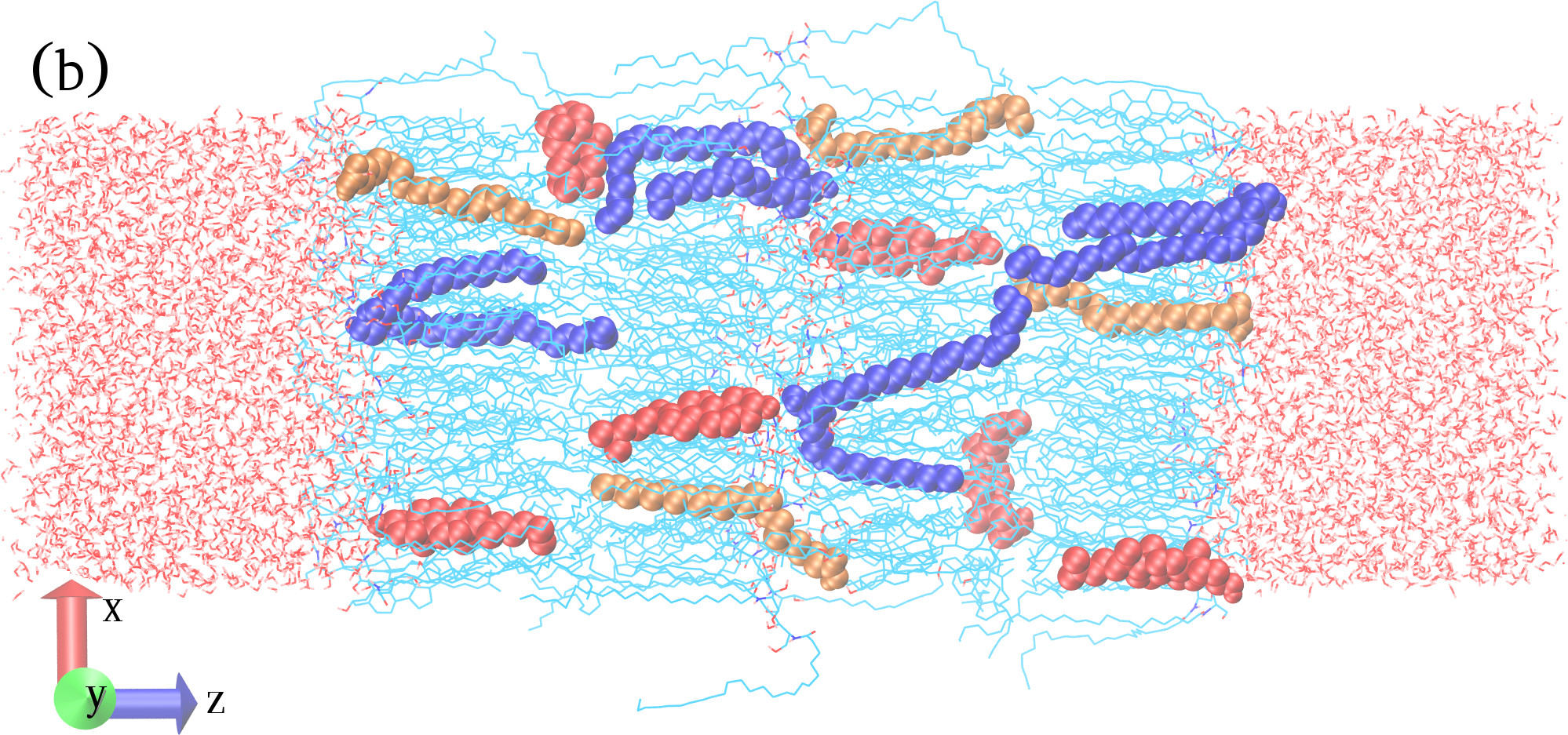}}
\centerline{\includegraphics[width=\linewidth,clip=]{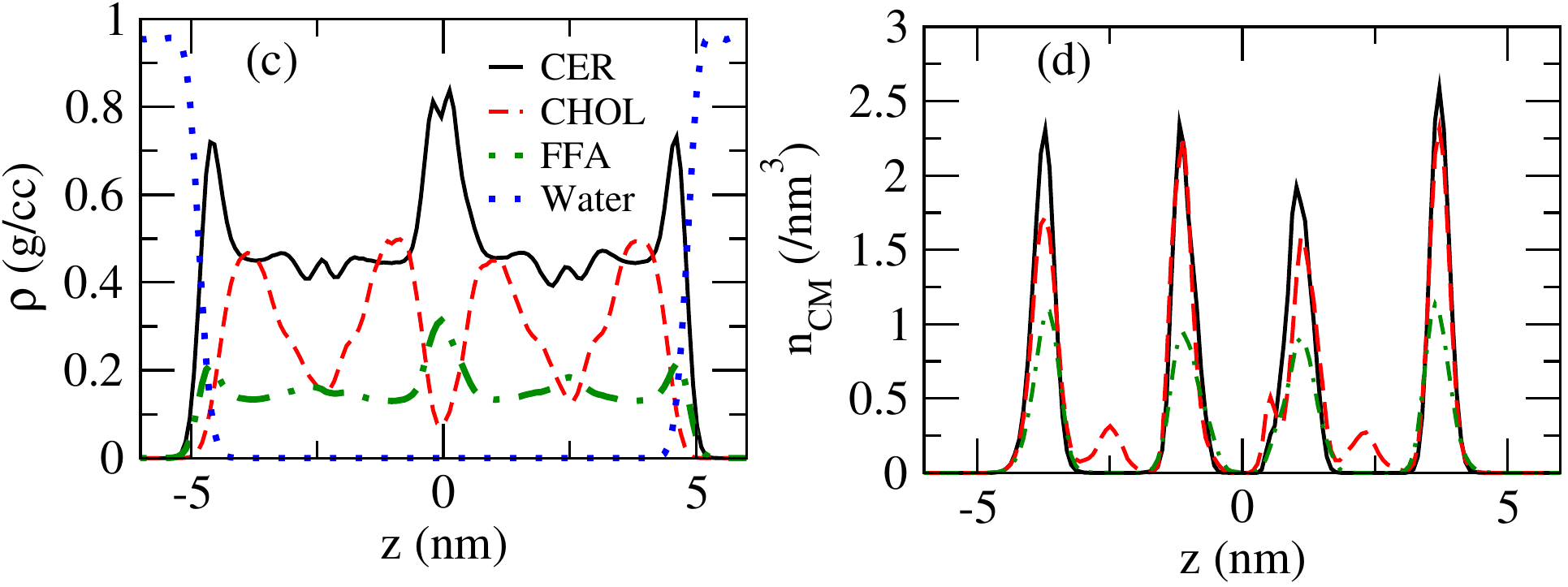}}
\caption{ (color online). (a) Schematic representation of the lipid molecules showing
 the polar atoms. (b)  Simulation snapshot with representative 
highlighted  lipids (red=CHOL, blue=CER; orange=FFA). The box dimension is 
$4.96 \times  5.03 \times  16.12\;\textrm{nm}^3$.  
(c) Mass densities of the lipids and water and (d) number density of center of mass of the lipids;  
CER (black solid line), CHOL (red dashed line), 
FFA (green dot-dashed line), water (blue dotted line).} 
\label{fig.denprof}
\end{figure}

Fully hydrated bilayers have frequently been simulated to
understand the lipid arrangements within and permeation through SC bilayers \cite{holtje.fachol.01,
pandit.cer2.06,notman.dmso.07, das.bpj.09, das.smat.10}, and multilayer stability has been simulated under 
limited hydration \cite{das2013lamellar,engelbrecht.smat.11}. 
An important result is that  the different tail lengths intrinsic to SC lipids (Fig.~\ref{fig.denprof}a) lead to a 
unique sandwich structure with a disordered liquid-like region between leaflet tails (Fig.~\ref{fig.denprof}b)  
\cite{das.bpj.09,das.smat.10,das2013lamellar}.

Here 
we consider two bilayers in excess water but without interbilayer water,  
comprising the ceramide N-lignoceroyl-D-erythro-sphingosine (CER NS 24:0),
 lignoceric acid (a common free fatty acid, FFA~24:0), and cholesterol (CHOL)  (Fig.~\ref{fig.denprof}a).  
We calculate the free-energy
of swelling and find a large ($\sim 3.6 \,k_{\textrm{B}} \textrm{T}$/water molecule) barrier for initial water ingress,  
suggesting that the analysis of SC lipid matrix function should invoke the properties of dehydrated SC lipid multilayers.  From long ($1 \,\mu \textrm{s}$)
molecular dynamics simulations, we find that strong hydrogen bonds between the head-groups of apposing
leaflets (from adjacent bilayers) force these two leaflets to move together as an 
{\em inverse bilayer}, with   
the hydrophobic tails in the disordered regions sliding against each other.
 CHOL  in these disordered interleaflet regions is in dynamic equilibrium with CHOL in the 
 ordered (gel) part  of the 
leaflets, which facilitates  transfer of CHOL between leaflets  that is orders of magnitude faster than in fluid phospholipid bilayers. 
This fast flip-flop can reduce
the curvature elastic cost in bending of the lipid multilayers
by redistributing CHOL asymmetrically between the two leaflets, thus enhancing the pliability and energy absorbing effects of the SC and hence skin.

\noindent{\bf Simulations:}
We use the `Berger' force field \cite{ryckaert.ff.75, jorgensen.ff.88, chiu.ff.95, berger.ff.97}
to describe the lipid interactions and
the SPC model \cite{spcwater} for the water molecules. 
The topology and the partial charges used for the lipid
molecules have been reported elsewhere 
\cite{Supplementary,das.bpj.09}.
Molecular dynamics 	simulations at 
constant temperature (340~K) and pressure (1~atm)
were carried out with GROMACS molecular 
dynamics software \cite{gromacs95, *gromacs05} using 
Nos\'e-Hoover thermostats separately coupled to the
lipids and the water molecules, and with a 
Parrinello-Rahman barostat.
Standard periodic
boundary conditions were applied in all three directions. 
Bond lengths were constrained with the LINCS algorithm for
the lipid molecules and the SETTLE algorithm for the water
molecules.  Long-range electrostatics contributions were
calculated with Particle Mesh Ewald summation (PME) and
a cut-off of 1.2~nm was used for both the Lennard-Jones
and the short-range electrostatics interactions. The timestep was 2\,fs.

As the starting configuration we use an 
equilibrated hydrated bilayer from an earlier study \cite{das.bpj.09}. 
The water molecules were removed and the lipid bilayer was 
repeated once along the bilayer-normal (z-direction) and was then rehydrated
with 5250 water molecules. The double bilayer contained 
112 CER, 112 CHOL and 64 FFA molecules.
After density equilibration, the configuration
was evolved for a further $1\,\mu\textrm{s}$. Configurations
were stored at $0.2\,\textrm{ns}$ intervals.

From the final configuration of this multilayer system,
we randomly select one test water molecule and pull it through 
the lipid multilayer 
at a speed of $0.05\,\textrm{nm/ps}$ along the z-direction. Configurations 
were stored every $0.2\,\textrm{nm}$. These saved
configurations were then evolved with the test molecule 
constrained to have the same $z$-separation from the lipid center of mass,
for longer than the relevant force-autocorrelation times ($2-40\,\textrm{ns}$).
The average force $\tfav$ on the constrained water molecule can then
be used to calculate 
the excess chemical potential \cite{marrink.jpc.94}. 
The whole procedure
was repeated for six different water molecules to calculate
statistical errors. 

\noindent{\bf Lipid structure and dynamics:} Fig.~\ref{fig.denprof}b shows a typical simulation snapshot
with a few highlighted lipid molecules. 
CER and FFA show high nematic order close to the head groups
and a disordered environment in the tail-tail interface of each bilayer (Fig S3, Supplementary Information \cite{Supplementary}). Some of the CHOL 
occupy this liquid-like tail-tail interface region. The head-head contact region ($z=0$)
shows large mass densities
from the CER and FFA (Fig.~\ref{fig.denprof}c), signaling
better alignment of the lipids when a leaflets is in contact with one
from an adjacent bilayer ($z=0$)
as opposed to when they are in contact with water ($z=\pm 4.5 \textrm{nm}$). 
Fig.~\ref{fig.denprof}d shows  that the CER and FFA centers of mass are sharply localized
at the centers of the leaflets, while  the CHOL center of mass has a subpopulation
($\simeq 8.3 \%$ or $\simeq9.3$ molecules) in the tail-tail interface. 

\begin{table}
\begin{ruledtabular}
\begin{tabular}{clcccc}
&   lipid          & inter HB & intra HB & H$_2$O HB &total \\\hline
outer & CER & -- & 0.94  & 2.02 & 2.96\\
leaflet& CHOL & -- &  0.26 & 0.58 & 0.84\\ 
& FFA  & -- &  0.36 & 1.30 & 1.66\\\hline
inner & CER &  0.70  & 1.27   & -- & 1.97\\
leaflet &CHOL  &  0.29  & 0.36   & -- &0.65\\
 &FFA  &  0.70  & 0.72   & -- &1.42\\
\end{tabular}
\end{ruledtabular}
\caption{Hydrogen bonds (HBs) per molecule between a given  inner  or outer leaflet lipid, 
and   lipids on either the same  (intra) or different (inter) leaflets, or with water. Inner-leaflet 
lipids have fewer total HBs, but a significant fraction (42\%) of them are inter-leaflet HBs.}
\label{tab.Hbond}
\end{table}

There are a large number of both intra-leaflet and inter-leaflet (and thus inter-bilayer) hydrogen bonds (Table~\ref{tab.Hbond}). There are more total hydrogen bonds per lipid on 
the outer leaflets, presumably because of the greater flexibility afforded by the solvent degrees of freedom.  While the outer-leaflet lipids hydrogen bond with water, the corresponding inner-leaflet lipids can replace some of these solvent hydrogen bonds by inter-leaflet hydrogen bonds, 
which effectively glue the inner leaflets together and make the dynamics of a 
double bilayer very different  from a single hydrated bilayer.

\begin{figure}[hpbt]
\centerline{\includegraphics[width=\linewidth]{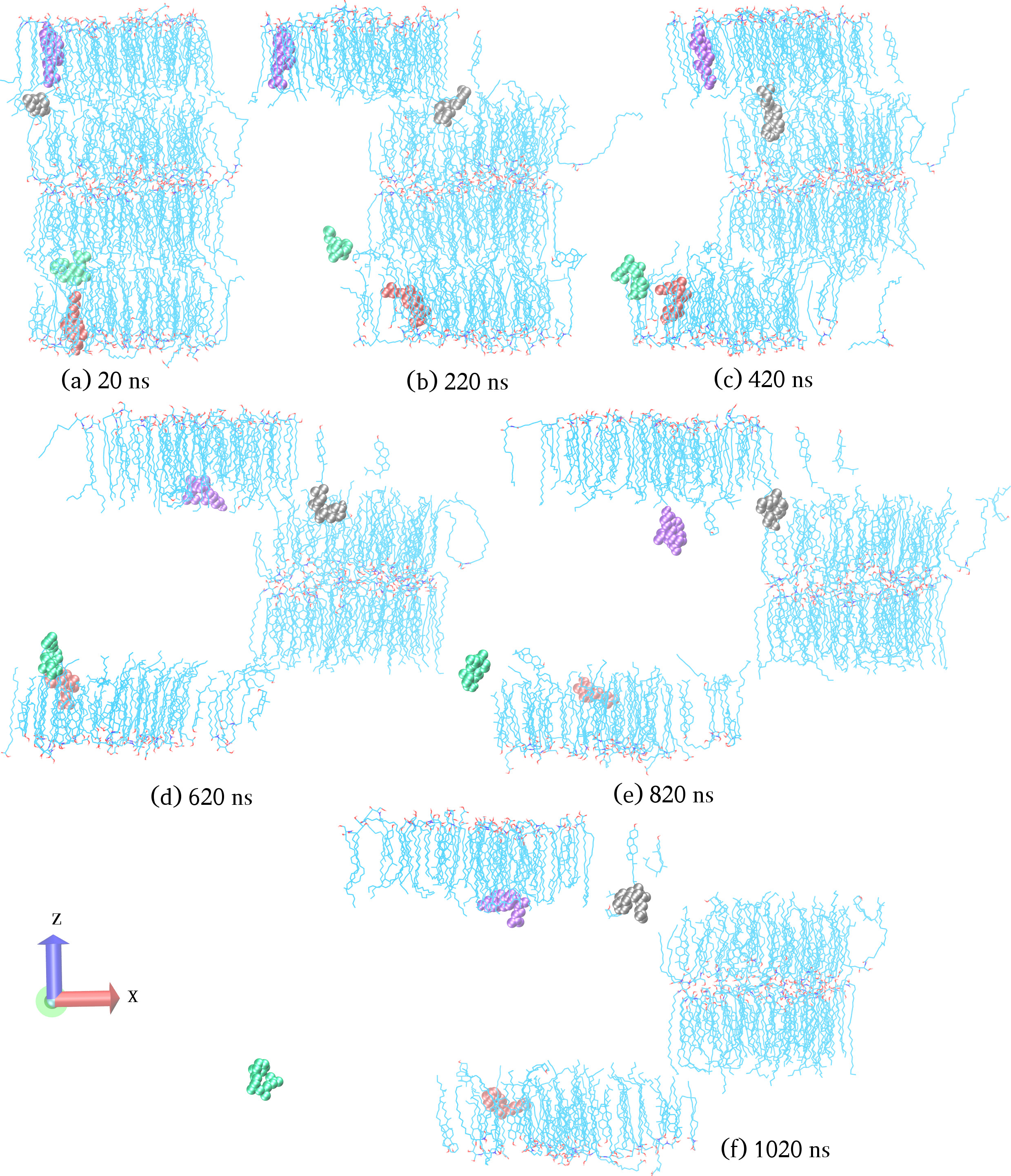}}
\caption{ (color online). Snapshots of the lipid molecules in the lipid
center of mass frame. For clarity, only lipid molecules with $y$-component of the initial center of mass less than 3.5 nm are shown. Four of the CHOL molecules are highlighted with spheres.
The inner two leaflets move together because of the inter-leaflet 
hydrogen bonding, and CHOL exchanges frequently between the ordered region and
the subpopulation between leaflets, leading to rapid flip-flop.
An animation of the trajectory is in the online supplementary material \cite{Supplementary}.}
\label{fig.flipflop}
\end{figure}

Fig.~\ref{fig.flipflop} shows snapshots during the $1\,\mu\textrm{s}$ trajectory. 
The two inner leaflets diffuse together 
coherently in the center of mass frame of the lipids.
Thus, a  stack of SC lipid multilayers 
behaves like a collection of `inverse bilayers'  wherein two leaflets 
belonging to adjacent traditional bilayers are strongly coupled by the head 
groups and slide relatively
easily at the hydrocarbon tail-tail interface. This is
the opposite from hydrated phospholipid  biological membranes.

In Fig~\ref{fig.flipflop} we have highlighted four
CHOL molecules,  showing  that CHOL frequently
exchanges between the
ordered regions of leaflets and the liquid-like tail-tail inter-leaflet region. 
SC lipids \textit{in vivo} are in a glassy or gel state; despite local segmental 
motion (\textit{e.g.} slithering of the tails), 	
two dimensional diffusion only occurs through slow cage-hopping \cite{das.bpj.09}.
However, the CHOL molecules in the liquid-like region
can readily diffuse, both translationally and rotationally, which allows
high overall CHOL mobility.  A molecule
in the ordered region can transfer to the liquid-like region and move
a large distance. Because of the rapid tumbling and head-group reorientation due to rotational diffusion, it can easily be  reabsorbed into  either of the leaflets.
The vacancy left when CHOL exits the ordered regions in turn
increases the in-plane mobility of CER and FFA molecules.

From the peaks of CHOL center of mass density $n_{CM}^{\textrm{CHOL}}$
(Fig.~\ref{fig.denprof}d) we identify zones
associated with the ordered and liquid-like regions of the bilayers, with boundaries specified 
by the local minima of $n_{CM}^{\textrm{CHOL}}$. 
We define a transition between zones when the center of mass of a molecule
penetrates  10\%  of the way into a neighboring zone, and remains in the new zone for at least 
0.4~ns. In the $1\;\mu\textrm{s}$ long
trajectory, we identified 322 such transitions involving
35 individual molecules (31\% of total CHOL content), while 6 molecules (5\% of CHOL) occupied
both leaflets at different times. Assuming steady state, the  characteristic  time scale for
CHOL exchange from an ordered leaflet to the central liquid-like zone 
is $\sim 0.64 \mu s$, and the characteristic \textit{flip-flop time}  
for CHOL exchange between the two leaflets is $\tau_{\textit{ff}}\simeq19 \mu s$ \footnote{In steady state, detailed balance implies that  $N_{21}\equiv n_1k_{21}\,dt = N_{12}$, where $n_i$ is the number in a given zone,  $k_{ji}$ is the rate, per molecule, that a molecule in zone $i$ transits to zone $j$, and $N_{ji}$ is the total number that transit in a time $dt$. Hence, if there are $N$ total transitions between two zones in time $T$, then the characteristic time $\tau_{1\rightarrow2}=k_{21}^{-1} =2 n_1 T/N$. 
For $N=322$ flip events in $T=1\mu\textrm{s}$, 
$n_{\textrm{\tiny O}} \simeq 102.7$ and
$n_{\textrm{\tiny D}} \simeq 9.3$, giving
$\tau_{\textrm{\tiny O} \rightarrow \textrm{\tiny D}} = (2 \times 102.7/322) \mu\textrm{s} = 0.64 \mu \textrm{s}$ and 
$\tau_{\textrm{\tiny D} \rightarrow \textrm{\tiny O}} =0.06 \mu \textrm{s}$.
For flip-flop events we consider all 112 CHOL molecules to be members of one of the two leaflets (exploring an ordered and a disordered zone) until the flip-flop event takes place (to another ordered zone).  Hence,  6 flip-flop events in $1\mu\textrm{s}$ leads to  $\tau_{\textit{ff}} = 2\times (112/2)/6 = 18.7 \mu \textrm{s}$.
}.
Accounting for the inner and outer leaflets separately
leads to a CHOL exchange timescales from the ordered to disordered
region of $\sim 0.4 \mu s$ for the outer (hydrated) leaflets
and $\sim 1.2 \mu s$ for the inner leaflets.
The much faster CHOL exchange from the outer hydrated leaflets is
reflected in a much larger mean-square in-plane displacement 
for  all lipids in the outer leaflets, as compared to their counterparts in the inner leaflets (Fig.~S7, Supplementary Information \cite{Supplementary}).

\begin{figure}[hptb]
\centerline{\includegraphics[width=\linewidth, clip=]{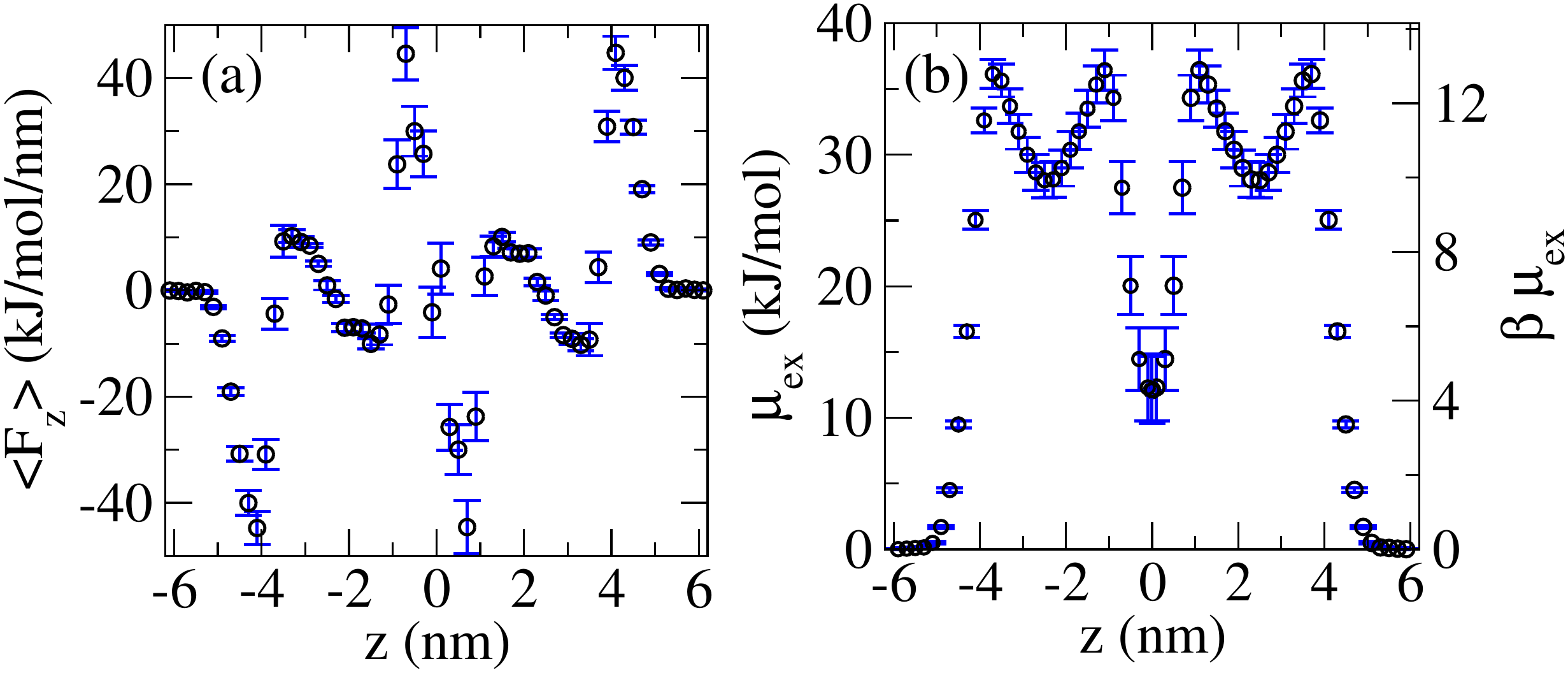}} 
\caption{(color online). (a) Average force $\tfav$ on a water molecule constrained at
a given $z$ from the lipid center of mass (averaged over 8 water molecules). 
(b) Excess chemical potential of water molecules. }
\label{fig.forceav}
\end{figure}

\noindent{\bf Excess chemical potential of water:}
From simulations with a water
molecule constrained at a given height  $z$, we calculate
the average $z$-component of the force $\tfav$ on the constrained water. 
Fig.~\ref{fig.forceav}(a) shows $\tfav$ averaged over time and six different water molecules. Because of the symmetry about the lipid center
of mass, we expect $\tfav$ to be an odd function of $z$. We exploit this
symmetry in calculating the excess chemical potential, 
$\mu_{ex} (z) = - \int_{\infty}^z  \langle F_{z'} \rangle dz'$,
from numerical integration of $\tfav$ (Fig~\ref{fig.forceav}(b)). The maximum in the
ordered region is similar to that found in simulations of
hydrated SC lipid bilayers \cite{das.smat.10}. 
The excess chemical potential remains positive and large compared to 
the thermal energy ($10.3\pm3.6\,\textrm{kJ/mol} \simeq 3.6\pm1.2 k_B T$) at the 
bilayer-bilayer interface ($z=0$), which demonstrates that at equilibrium the SC multilayers do not undergo swelling.

\noindent{\bf Discussion:}
The long  ($1~\mu\textrm{s}$) simulations show that 
CHOL has a dynamic subpopulation at the disordered bilayer midplane,
which rapidly exchanges with CHOL in the ordered lipid
region on $\mu$s timescales. The 
 high rotational diffusivity (tumbling) of CHOL in the mid-leaflet
liquid-like region allows molecules from
the subpopulation to readily incorporate into the ordered region of
either of the adjacent leaflets, leading to fast flip-flop
times $\tau_{\textit{ff}}\simeq19\,\mu\textrm{s}$.

CHOL flip-flop has been investigated extensively for phospholipid bilayers, yielding much slower timescale than found here.
Different experiments assign a timescale of milliseconds to hours 
\cite{Garg.BPJ.11, Bruckner.BPJ.09}. 
From potential of mean force calculations using an all atom DPPC bilayer with
40\% CHOL concentration, the flip-flop time scale of CHOL was estimated
to be $\sim 50~\textrm{ms}$ \cite{Bennet.jacs.09}. Similar time scales were estimated from simulations of bilayers containing 1:1:1 molar ratios of palmitoylsphingomyelin (PSM), 1-palmitoyl-2-oleoyl-sn-glycero-3-phosphocholin (POPC), and CHOL \cite{bennett.JLR.12}. 
Naively one would assign a much smaller flip-flop rate for
CER bilayers, in which the small ceramide headgroup and prevalence of single chain fatty acids lead to relatively high densities in the ordered
region, which would imply much lower molecular mobility, particularly in the gel phase found \textit{in vivo} (and in our simulations). 

However, CHOL flip-flop is enhanced by a number of effects that are specific to the SC 
membrane: (1)  strong hydrogen bonding among CER molecules leads to an ordered dense
leaflet (Fig.~\ref{fig.denprof}), which is comparatively less favorable for CHOL than in phospholipid solid (or 
fluid) phases; (2) the length asymmetry of the CER tails leads to a  low-density
liquid-like inter-leaflet region, within which the CHOL can reside (Fig S3, Supplementary Information \cite{Supplementary}); (3) 
the higher free-energy of CHOL in the ordered region lowers the barrier
for hopping into the  liquid-like region; (4) the relative disorder in the liquid-like region allows the 
CHOL to easily reorient to incorporate into another leaflet (Fig S3, Supplementary Information \cite{Supplementary}).

A flat multicomponent lipid bilayer is governed by a free energy $G$ that includes the asymmetry  $\delta\phi$ of the CHOL content between leaflets \cite{BenShaul.book.95,Bruckner.BPJ.09}:
\begin{equation}
G=\frac12\int d^2r\left[\kappa C^2 + 2\alpha C \delta\phi +\chi\delta\phi^2\right],
\end{equation}
where $C$ is the bending curvature,  $\kappa$ the bending modulus, $\chi$ the penalty for creating asymmetry, and the bending-asymmetry coupling $\alpha$ depends on the shape and energetics of CHOL packing into the leaflets. An imposed curvature $C$ can induce a CHOL fluctuation $\delta\phi\simeq-\alpha C/\chi$  on timescales longer than the 
flip-flop timescale $\tau_{\textit{ff}}$, which leads to a reduction of the bending modulus 
$\kappa \rightarrow\kappa_R=\kappa-\alpha^2/\chi$.
Thus, SC membranes can quickly 
adapt to widely different curvatures without the high elastic penalty expected for a gel-like 
phase. Such dynamic curvature changes are expected  {\em in~vivo}
due to hydration- and dehydration-induced shape changes of the corneocytes, and normal 
folding and stretching of skin due to physiological activities. 
With the low permeability of skin such hydration-induced
changes of corneocyte shape will happen much more slowly 
than flip-flop, which can act to keep the SC multilayers
free of curvature stress.

The subpopulation of CHOL within the disordered center region allows for anomalous in-plane diffusion controlled by adsorption-desorption between the ordered and disordered 
regions  \cite{BychukPrl95}, which is much faster than permitted in the dense lipid lamellae.
In turn, the CHOL dynamics enhances the mobility of other lipid species by creating temporary 
free volumes in the ordered leaflet. Such enhanced molecular mobility
should render skin more dissipative than similarly
packed long chain molecules. Hence,  flip-flop may be the dominant mechanism for 
 the experimentally-observed  enhancement of 
fluidity of SC lipids due to CHOL \cite{kitson.biochem.94}.

\noindent{\bf Summary:} 
We have carried out large-scale molecular dynamics simulations of stratum corneum lipid bilayers. The membranes are strongly dehydrated, with a barrier for aqueous swelling of multilayers that is governed by inter-leaflet hydrogen bonding. Hence, theories developed to describe  fully-hydrated phospholipids must be applied to SC multilayers with care. The gel-like phase found at physiological temperatures has a liquid-like disordered layer between leaflets, which facilitates rapid  cholesterol flip-flop and can significantly soften the bending modulus, as well as inducing mechanisms for greater dissipation. One expects a strong modulus with little dissipation at high frequencies $\omega>\tau_{\textit{ff}}^{-1}$, and a softer response and greater dissipation at lower frequencies.  These  effects are important for adaptation of the skin to changing conditions, as well as contributing to the skin's effective and remarkable resilience. 
The extensive hydrogen bonding within SC multilayers has some similarities with novel 
self-healing materials invented by Liebler {\em et~al.} \cite{cordier2008self}. In both cases, hydrogen bonds form and reform to control the mechanical properties and response of materials. Implementing an analog of CHOL flip-flop in self-healing materials might impart stronger dissipation and frequency dependent bending response.

The primary function of SC lipids is as a hydration barrier. Simulations
show \cite{das.smat.10} that CER alone provides orders of magnitude lower permeability
for water, when compared to three component CER:CHOL:FFA bilayer.  
The presence of CHOL decreases the tail order and thus increases permeability; however, CHOL also helps to soften the mechanical properties, both the intrinsic bilayer compressibility \cite{das.smat.10} and the bending modulus. Hence, Nature may have optimized the lipid composition so that the SC lipid matrix can be deformed rapidly and relatively easily, while still maintaining an acceptable hydration barrier.

This work was supported by Yorkshire Forward (YFRID Award B/302) and part financed by the European Regional Development Fund (ERDF). Computational resources were provided by SoftComp EU Network of Excellence. We gratefully acknowledge discussions with Anna Akinsina, Johan Mattsson, and Patrick Warren. 

\bibliographystyle{apsrev4-1}

%


\clearpage
\leftline{\bf Supplementary material:}
\vskip0.2cm
\leftline{\bf Fast cholesterol flip-flop and }
\leftline{\bf lack of swelling in skin lipid multilayers}
\vskip0.2cm
\leftline{ Chinmay Das, Massimo G. Noro and Peter D. Olmsted}

\setcounter{section}{0}
\setcounter{figure}{0}
\setcounter{table}{0}
\setcounter{equation}{0}
\renewcommand{\thesection}{S\arabic{section}}
\renewcommand{\thefigure}{S\arabic{figure}}
\renewcommand{\thetable}{S\arabic{table}}
\renewcommand{\theequation}{S\arabic{equation}}
\section{Analysis details and additional results}
\begin{figure}[htbp]
\centerline{\includegraphics[width=\linewidth]{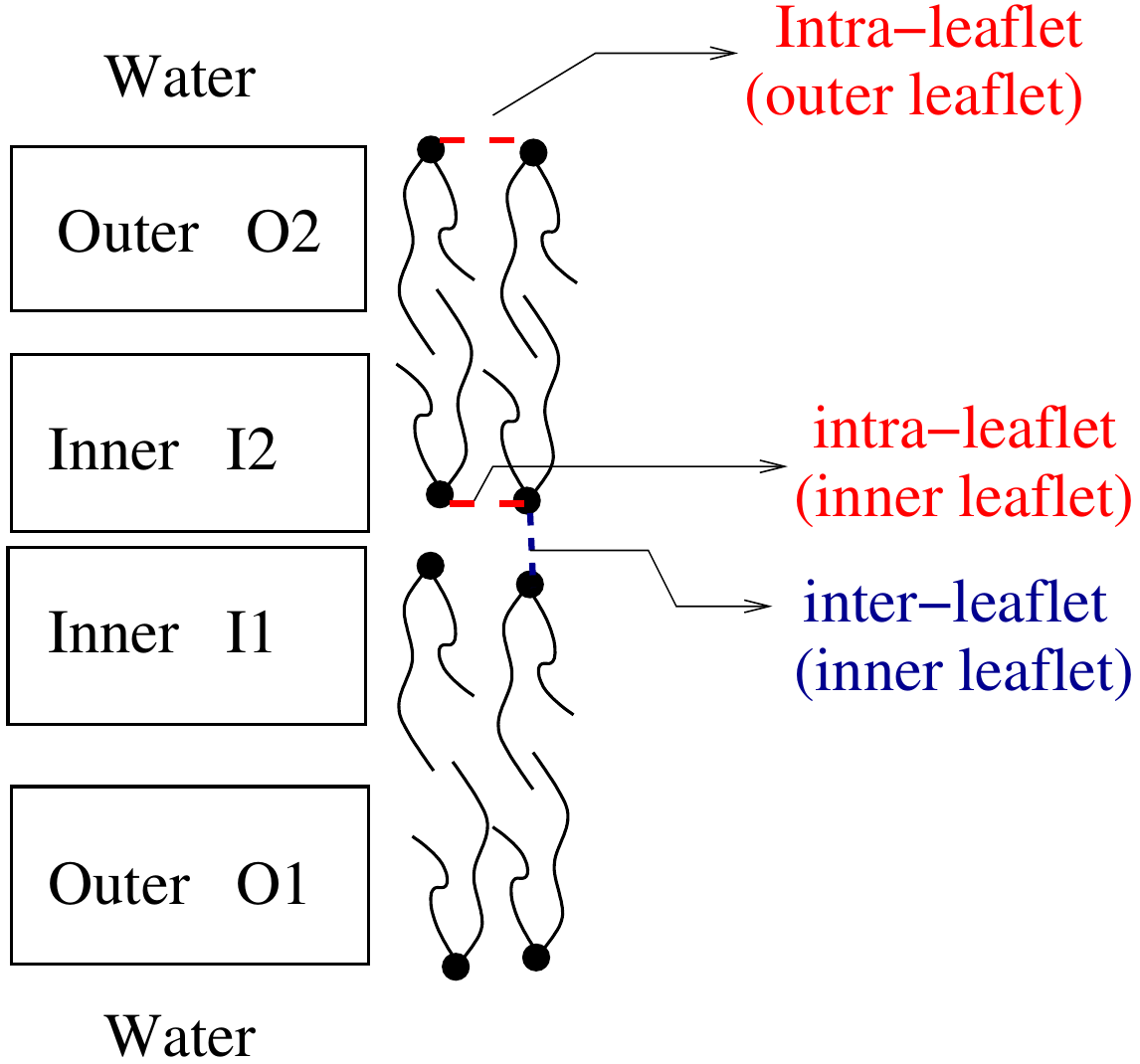}}
\caption{Naming convention for leaflets and inter-lipid hydrogen bonds.}
\label{fig.name}
\end{figure}
In this supplementary material, we provide the methods used in
calculating the hydrogen bonds, flip-flop time-scales, in-plane
molecular diffusion, and the excess chemical potential of water.
In our simulation of double bilayer in excess water, only two of the 
leaflets are in contact with water (Fig.~\ref{fig.name}). We term these
two leaflets as the `outer leaflets'. The other two `inner' leaflets face each
other and are not in contact with water. After an equilibration time
of 20~ns, we stored 5000 configurations separated by 0.2~ns spanning
a total of 1~$\mu$s. Unless otherwise stated, the results below are 
averaged over these 5000 configurations.

\subsection{Hydrogen bonds}
\begin{figure}[htbp]
\centerline{\includegraphics[width=5cm]{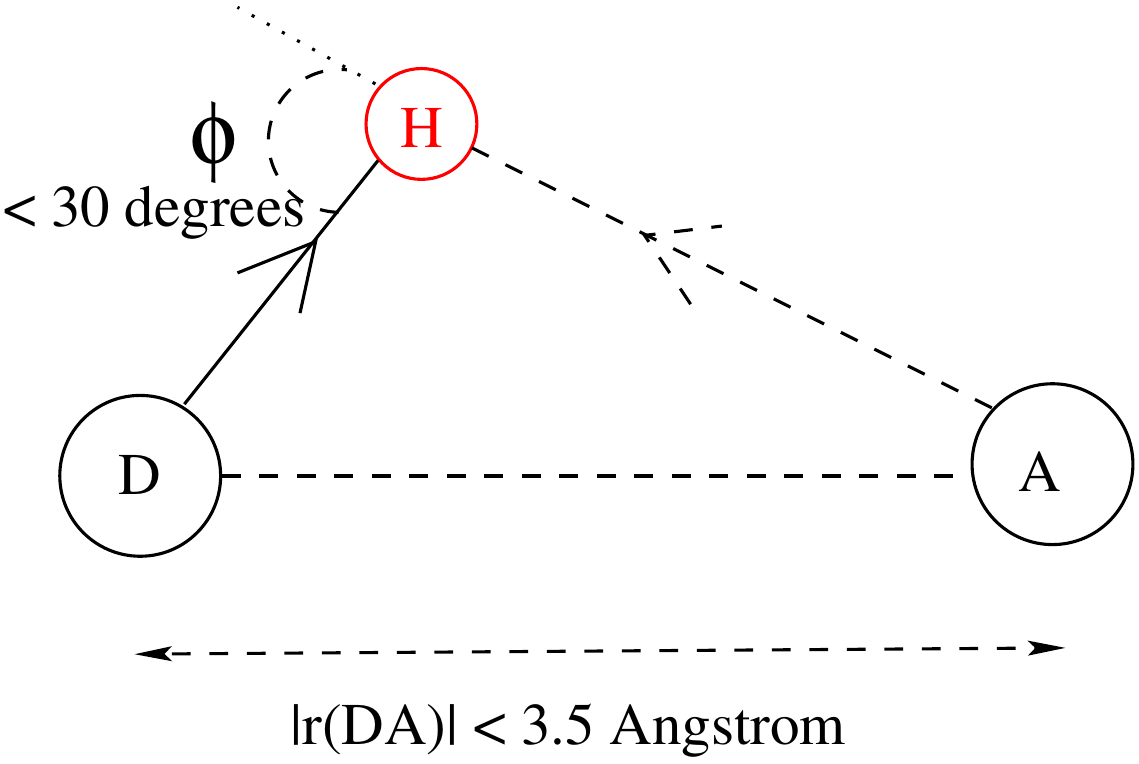}}
\caption{Geometric criteria used to identify hydrogen bonds.}
\label{fig.hbnddef}
\end{figure}

We use a geometric criteria \cite{ferrario.jcp.90, luzar.jcp.93, luzar.prl.96,
torshin.PE.02}  to define a hydrogen bond if the
distance between the donor and the acceptor atoms is less than
$3.5$\AA\ and simultaneously the absolute angle between the
vectors $\vec{r}_{DH}$ and $\vec{r}_{AH}$
is less than 30$^{\circ}$ (Fig.~\ref{fig.hbnddef}).

Table~\ref{tab.hbnd} shows the average number of lipid-lipid
hydrogen bonds for each lipid species. We have separated these into intra-leaflet and inter-leaflet (for inner leaflets that face each other) 
hydrogen bonds. The inner leaflet lipids form a
large number of intra-leaflet, as well as inter-leaflet, hydrogen bonds with lipids. In addition, lipids in the outer leaflets are also involved in
a large number of hydrogen bonds with water molecules.

\begin{table}[htbp]
\begin{ruledtabular}
\begin{tabular}{lcccccc}
Lipid & \multicolumn{3}{c}{Outer leaflets} & \multicolumn{3}{c}{Inner leaflets} \\ \cline{2-4} \cline{5-7}
& lipid & water & total & intra & inter & total \\ \hline \hline
CER & 0.94 & 2.02 & 2.96 & 1.27 & 0.70 & 1.97  \\  
CHOL& 0.26 & 0.58 & 0.84 & 0.36 & 0.29 & 0.65 \\
FFA & 0.36 & 1.30 & 1.66 & 0.72 & 0.70 & 1.42 
\end{tabular}
\caption{Number of  hydrogen bonds per lipid molecule in the different leaflets.}
\label{tab.hbnd}
\end{ruledtabular}
\end{table} 
\subsection{Tail order}

\begin{figure}[htbp]
\centerline{\includegraphics[width=\linewidth]{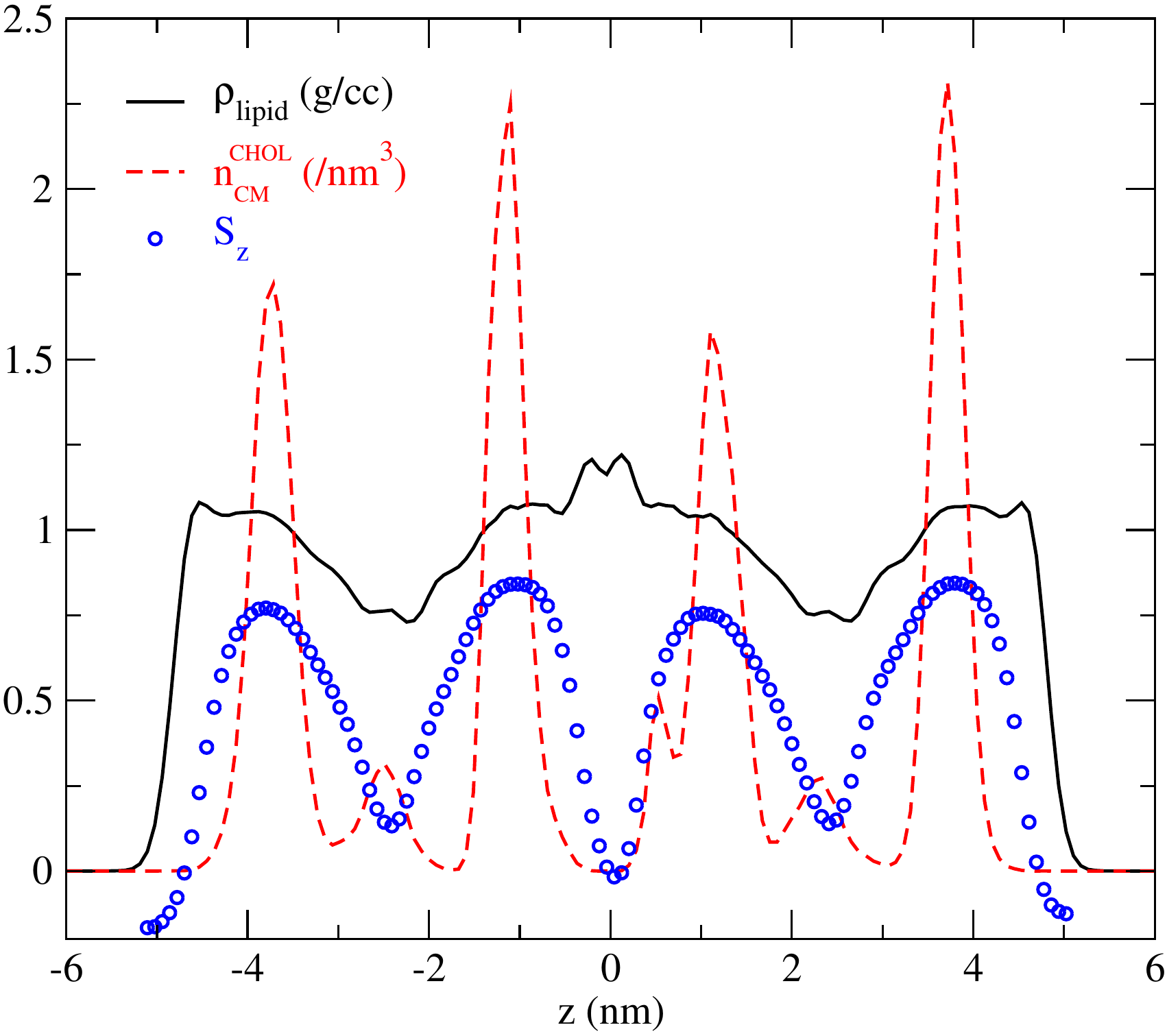}}
\caption{Total lipid mass density $\rho_{\textrm{lipid}}$ (black solid line), number density of CHOL center of mass
$n_{\textrm{CM}}^{\textrm{CHOL}}$ (red dashed lines) and tail order parameter $S_z$ (blue symbols) as a function of distance from lipid
center of mass.}
\label{fig.op}
\end{figure}

To investigate the alignment of the lipid tails,
for any three consecutive CH$_2$ groups (C$_{i-1}$, C$_{i}$ and C$_{i+1}$) in the CER or FFA molecules, we
consider the angle $\theta_z$ of the vector (C$_{i+1}$ - C$_{i-1}$) with respect to the z-axis (normal
direction to the lipid layers). We define an order parameter \cite{vermeer.EBJ.07}
\begin{equation}
S_z (z) = \left\langle
\frac{3 \cos^2 \theta_z - 1}{2}
          \right\rangle, 
\end{equation}
where the angular bracket denotes averaging over all CH$_2$ triplets with the central group being at 
a distance $z$ from the lipid center of mass. The usual order parameter is calculated (or measured) as a function of carbon number along the lipid tails \cite{vermeer.EBJ.07}, while we consider it as a function of
$z$. For perfect tail alignment along the $z$-direction, $S_z=1$; random order gives $S_z=0$; and perfect aligment perpendicular to the $z$-direction gives $S_z=-0.5$.
Fig.~\ref{fig.op} shows that $S_z$ becomes close to zero in the tail-tail interface region ($z \simeq \pm 2.4 \textrm{nm}$)
signifying a disordered region. The same region shows lower total mass density and a subpopulation of
CHOL center of mass.

\subsection{Density profile and CHOL flip-flop}
We unfold the molecules in the saved configurations and fix the lipid center of mass at the 
origin. The mass density of
the lipids and the number density of the center of masses of the lipids
were calculated in this lipid center of mass reference frame. 

\begin{figure}[htbp]
\centerline{\includegraphics[width=\linewidth]{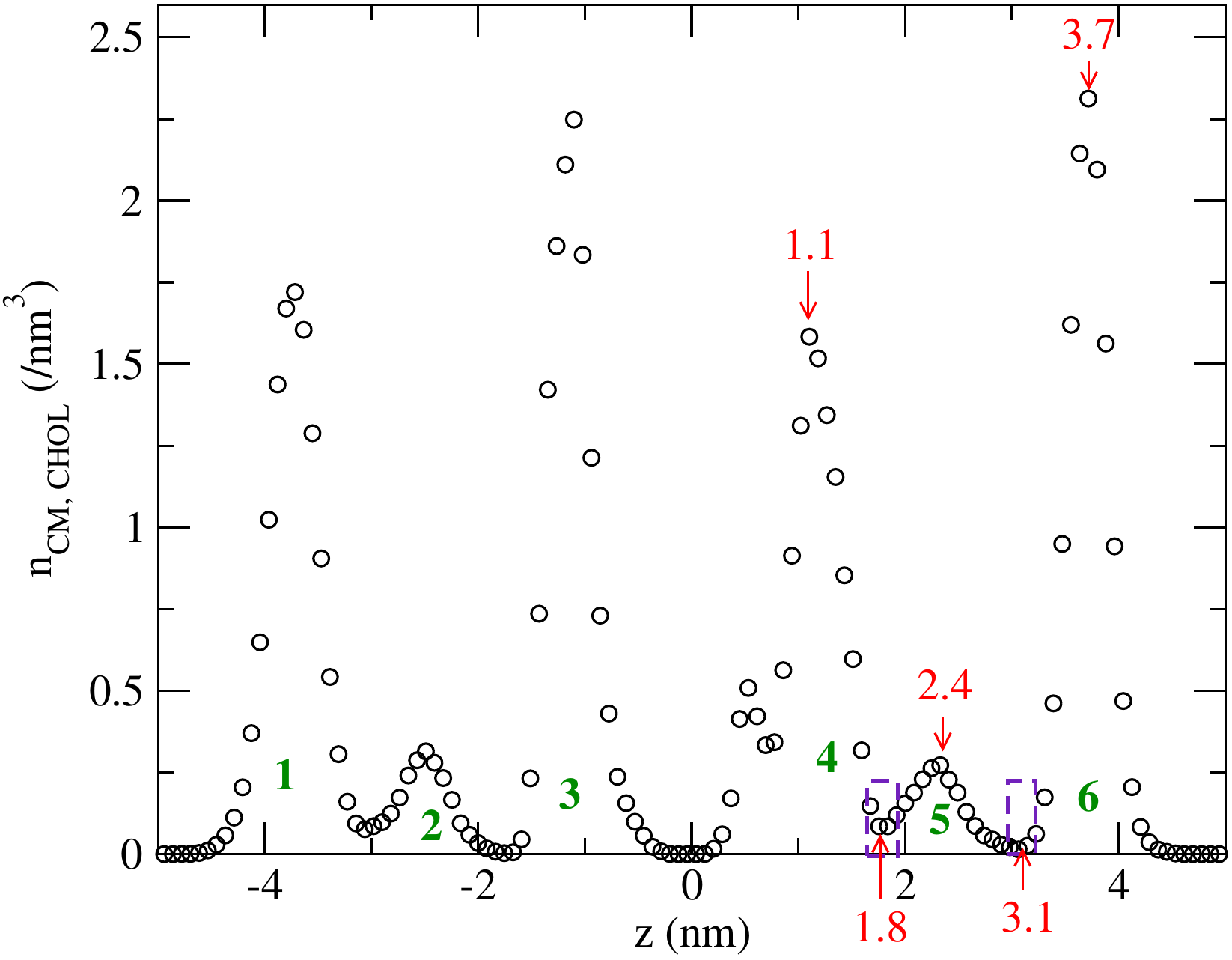}}
\caption{Average density of center of mass of CHOL and identification
of zones (1-6). The (red) arrows show the boundaries and local density maximum of zone 
5, while the (violet) dashed boxes show the criteria for a `flip', which is a transfer between 
adjacent zones.}
\label{fig.cholzone}
\end{figure}
From the distribution of
the centers of mass of CHOL, we identify two inter-leaflet liquid like
regions centered at $z=\pm 2.4\;\textrm{nm}$. 
For the peak at $z=2.4 \;\textrm{nm}$, the minima in the
distribution of CM are at $z=1.8\;\textrm{nm}$ and $z=3.1\;\textrm{nm}$.
In the first configuration, we assign a zone index to a given CHOL 
depending on the $z$-coordinate of its center of mass:
\begin{align}
\textrm{(ordered) zone 1:} &  &&  z_{\scriptscriptstyle CM} \le -3.1 \nonumber\\
\textrm{(disordered) zone 2:} && -3.1 \le\,\, & z_{\scriptscriptstyle CM} < -1.8\nonumber\\
\textrm{(ordered) zone 3:} &&  -1.8 \le\,\, & z_{\scriptscriptstyle CM} < 0 \nonumber\\
\textrm{(ordered) zone 4:} &&  0  \le\,\, & z_{\scriptscriptstyle CM} < 1.8 \nonumber\\
\textrm{(disordered) zone 5:} && 1.8 \le\,\, & z_{\scriptscriptstyle CM} < 3.1\nonumber\\
\textrm{(ordered) zone 6:} &&  3.1 \le\,\, & z_{\scriptscriptstyle CM} .\nonumber
\end{align}
In subsequent configurations
we assign a new provisional zone index to a given CHOL only if has moved into the next zone by at least 10\% of width of the next zone. This
criteria is indicated by dashed boxes in Fig.~\ref{fig.cholzone}.
Thus, a CHOL initially belonging to zone 5 is considered to have
moved to zone 4 only if its $z_{CM} < 1.64$.  

We define a  \textit{flip} event as  a transit between that does not revert to the original zone 
within two frames  (0.4ns). There are nearly 3 times more flip events involving
the outer leaflets than involving the inner leaflets.  We define a \textit{flip-flop} event to be when a CHOL  enters a ordered zone after having 
entered the disordered zone from a different ordered zone. In our trajectory
we did not find any lipid exchange between the bilayers, so that
all flip-flop events involve CHOL exchange between an inner leaflet and an outer leaflet.
 Table~\ref{tab.flip} shows the data used in calculating the flip-flop time scales. 

\begin{table*}[htbp]
\begin{ruledtabular}
\begin{tabular}{llcclccc}
Transition  & \multicolumn{2}{c}{Starting zone} & $n_{\textrm{start}}$ & \multicolumn{2}{c}{Finishing Zone}  & events & $\tau/\mu\textrm{s}$ \\ \cline{2-3} \cline{5-6}
& region & zone \# & region && zone \# && \\\hline
flip & outer ordered &(1, 6) &102.7& disordered &(2, 5) & 241 & 0.4 \\
 & inner ordered &(3, 4)  &102.7& disordered &(2, 5) & 81 & 1.2 \\
 & ordered &(1, 3, 4, 6) &102.7& disordered &(2, 5) & 322 & 0.64 \\
 & disordered &(2,5) & 9.3& ordered &(1,3,4,6) & 322 & 0.06 \\\hline
flip-flop & outer or inner leaflets  & (1,6) or (3,4) &112& inner or outer leaflets &(3,4) or (1,6) & 6 & 0.19
\end{tabular}
\end{ruledtabular}
\caption{Statistics for transitions of CHOL molecules between different regions in
 $1\,\mu\textrm{s}$, for calculations of flip times and flip-flop times. For flip-flop, CHOL begin in the outer (inner) ordered region of a leaflet, and explore the outer (inner) ordered and disordered regions until it enters the inner (outer) ordered regime of the other leaflet of the same bilayer.}
 \label{tab.flip}
\end{table*}

\subsection{Mean square displacement}
\begin{figure}[htbp]
\centerline{\includegraphics[width=\linewidth]{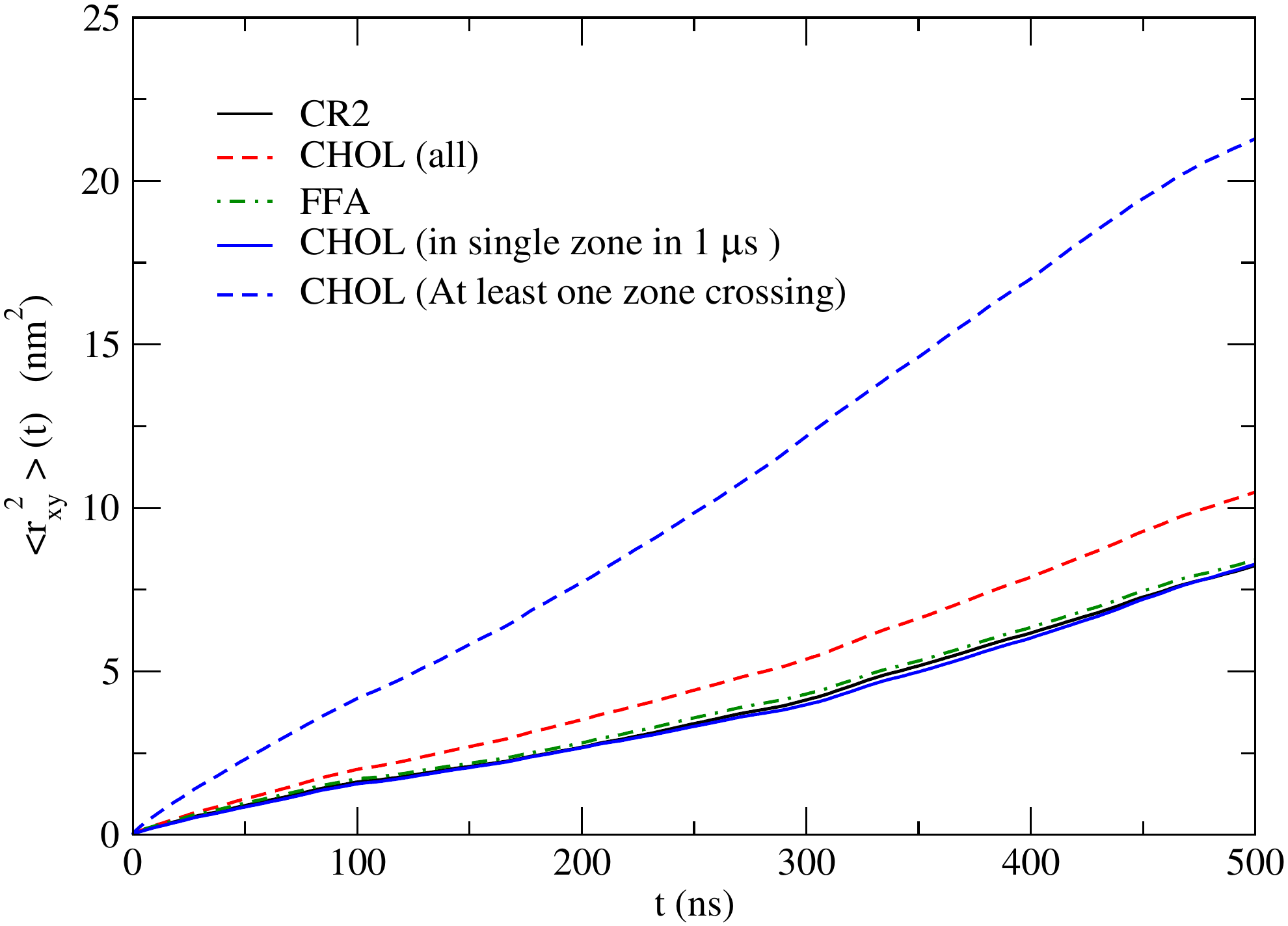}}
\caption{Mean square displacement of lipid center of mass in the
center of mass reference frame of all lipids.}
\label{fig.displipid}
\end{figure}
Fig.~\ref{fig.displipid}  shows the $x-y$ component of mean square
displacement of the center of mass of the lipids, in a reference frame
in which the center of mass of all the lipids is fixed. 
The data is averaged over the lipid molecules and over the time origin.
CHOL shows
higher in-plane mobility (red dashed line) than CER or FFA. We separate
 CHOL into populations that did and did not  undergo flip events during the entire trajectory.
 CHOL without any flip events show a mean-square in-plane displacement similar to that of CER or FFA, while CHOL with flip events have a much larger mobility: \textit{e.~g.}
 \begin{equation}
\left. \frac{\langle r_{xy}^2\rangle_{\textit{flip}}}{\langle r_{xy}^2\rangle_{\textit{no flip}}} 
\right|_{(\tau=500\,\textrm{ns})} \simeq 2.6.
\end{equation}

\begin{figure}[htbp]
\centerline{\includegraphics[width=\linewidth]{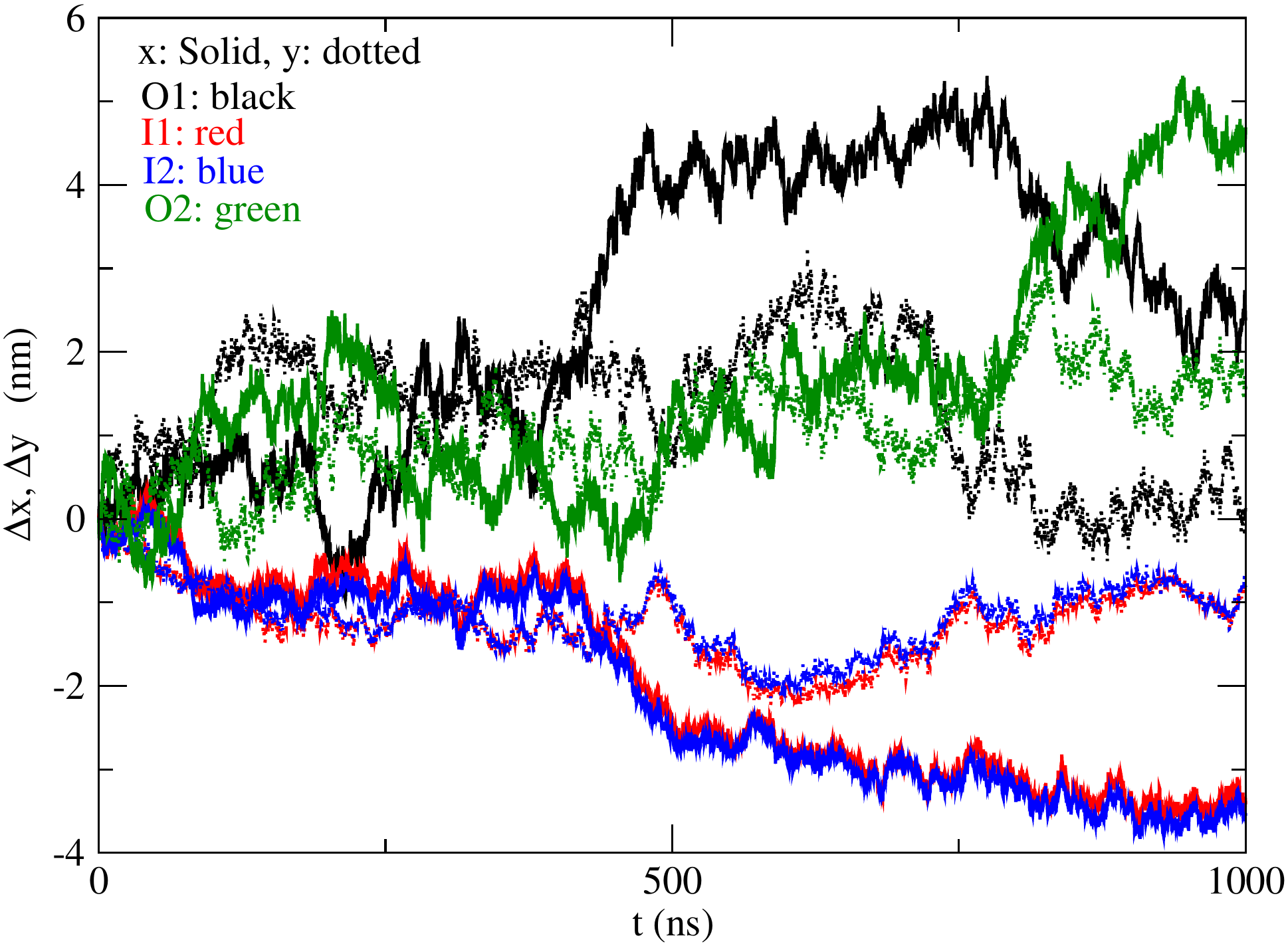}}
\caption{Diffusion of center of mass of the leaflets with respect to the center of
mass of all lipid molecules is fixed.}
\label{fig.displeaflet}
\end{figure}
In simulations of finite bilayers the leaflets can diffuse with respect to
each other \cite{denOtter.BPJ.07}. Since some of the CHOL molecules
move between leaflets,  we define an approximate leaflet center of
mass in terms of the CER and FFA molecules. Fig.~\ref{fig.displeaflet} 
 shows the diffusion of the leaflets' centers of mass. Both the
$x$ and $y$ components of the inner two leaflets (red and blue) move together
coherently over the entire trajectory. 

\begin{figure}[htbp]
\centerline{\includegraphics[width=\linewidth]{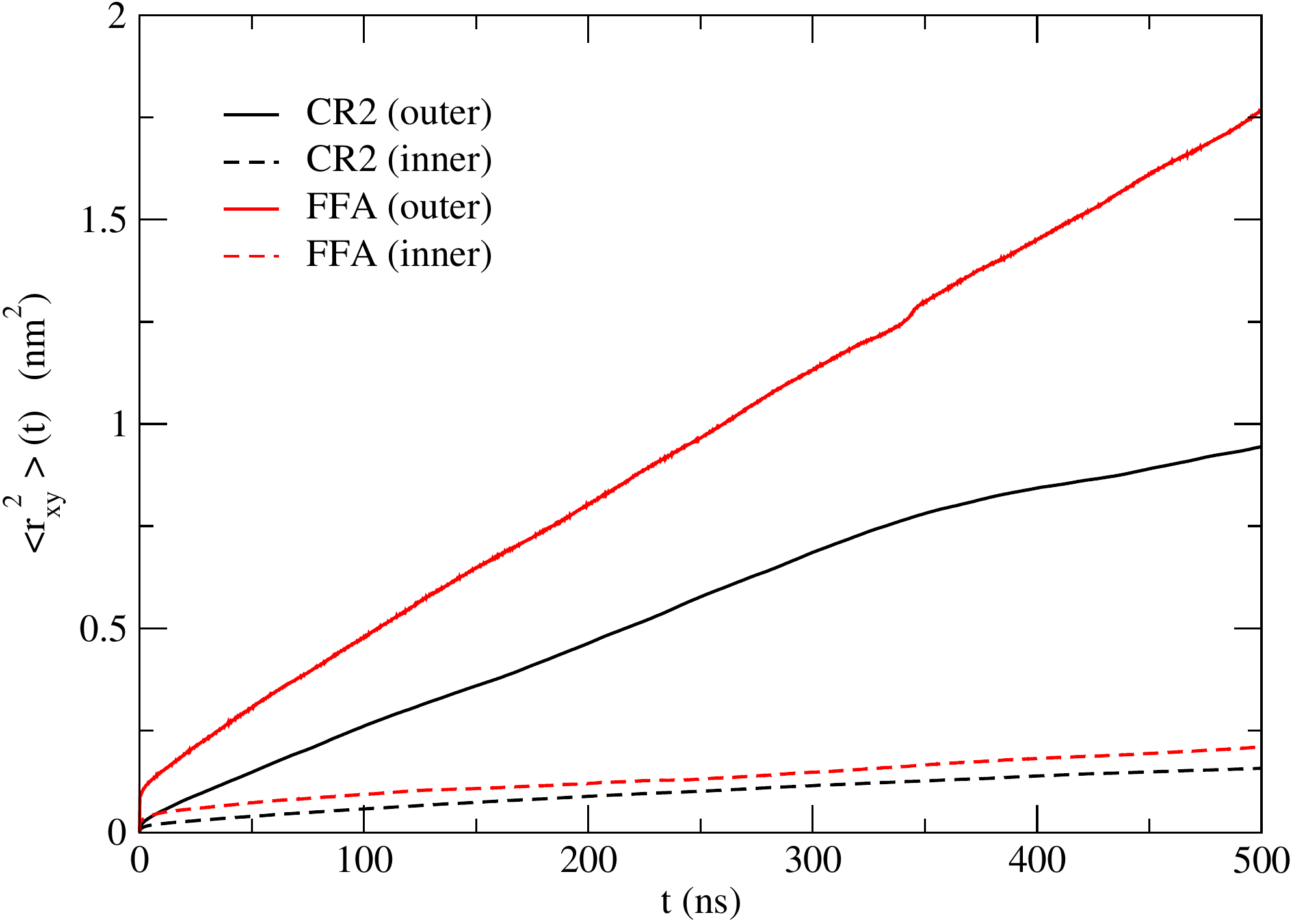}}
\caption{Mean square displacement of lipid center of mass in the reference frame of the leaflet
center of mass.}
\label{fig.dispinleaf}
\end{figure}
By calculating displacements in the reference frame of a given leaflet,
we obtain the mean square displacement of CER and FFA lipids in the reference frame
in which the leaflet center of mass is fixed, which is appropriate for a multilayer stack in which leaflet diffusion or sliding is prohibited. Fig.~\ref{fig.dispinleaf} shows that the lipids in the outer leaflets 
are significantly more mobile than those in the inner leaflets. Comparison of 
Figs.~\ref{fig.dispinleaf} and \ref{fig.displipid} shows that, even for the outer leaflets,
the main contribution to in-plane
displacement is from leaflet diffusion, which is pronounced here because of the small 
membrane size and periodic boundary conditions. The fast CHOL exchange and fewer 
inter-lipid hydrogen bonds render the lipids in the outer leaflets several times more mobile. 

\subsection{Force fluctuations for constrained water molecules}
\begin{figure}[htbp]
\centerline{\includegraphics[width=0.85\linewidth]{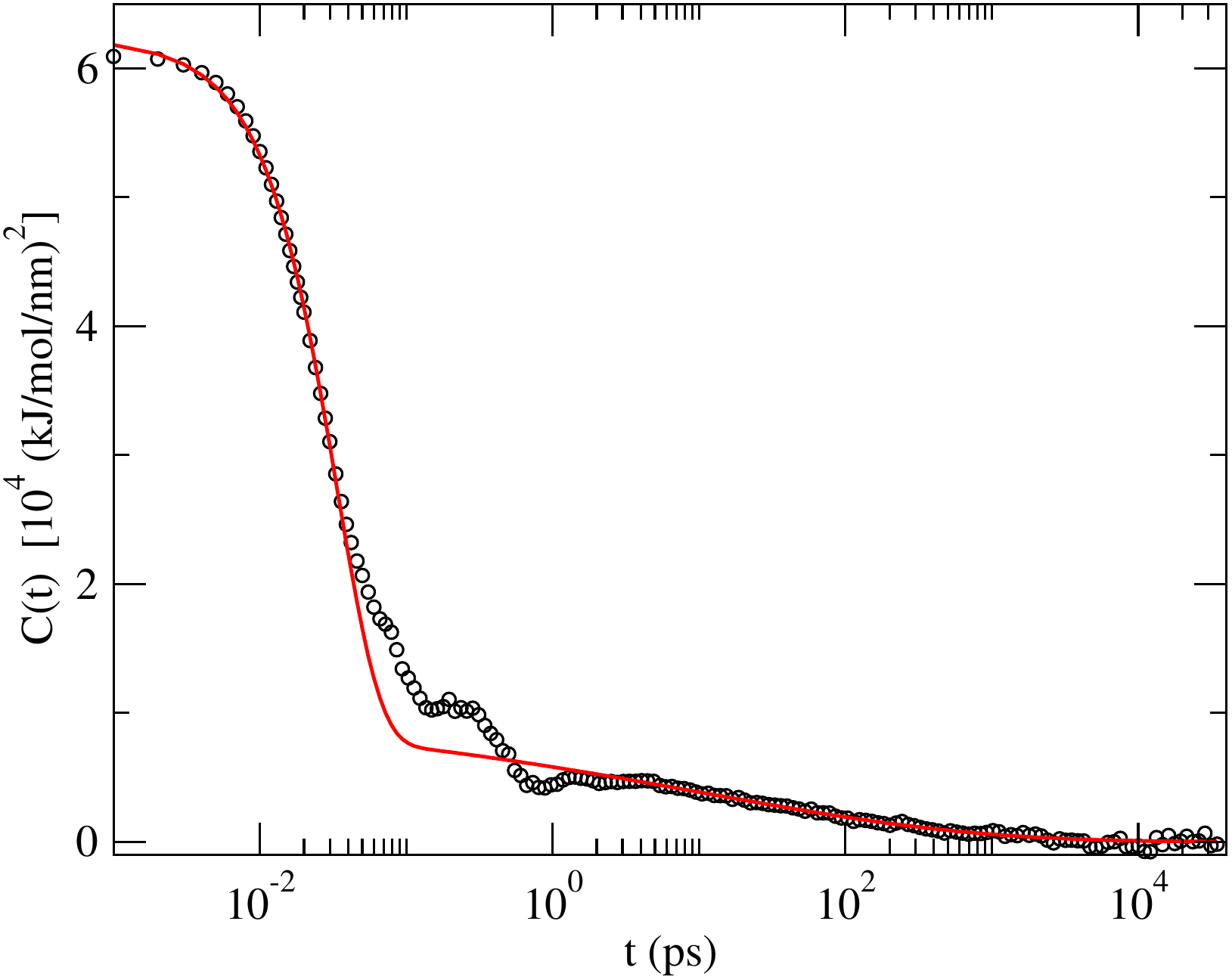}}
\caption{Autocorrelation of $F_z (z,t)$ (symbols) at $z=0.3\,\textrm{nm}$
with a fit (line) to Eq.~\ref{eq.fit}, with parameters $\tau_i=0.03\,\textrm{ps}$ and $7.9\,\textrm{ns}$ for the fast and slow times. The corresponding exponents are $\beta=1.3$ (fast: compressed exponential) and $\beta=0.18$ (slow: stretched exponential). 
}
\label{fig.fzcor}
\end{figure}

We perform simulations in which a single water molecule is constrained to
at a given $z$-separation from the lipid center of mass. This constraint induces a rapidly fluctuating force $F_z(z,t)$ in the $z$ direction.  However, we find a slowly decaying
time correlation in $F_z(z,t)$, particularly in the ordered regions.  At a fixed $z$, we fit the 
autocorrelation of $F_z(z,t)$  to a sum of two generalized exponentials:
\begin{align}
C(z,t) &\equiv \langle F_z(z,t) F_z(z,0)\rangle\\
& =\sum_{i=1}^2 A_i(z) \exp\left[- \left( \frac{t}{\tau_i(z)} \right)^{\beta_i(z)} \right],\label{eq.fit}
\end{align}
where the parameters $A_i, \tau_i$, and $\beta_i$ are fitted with the 
Levenberg-Marquardt damped least-square method. Fig.~\ref{fig.fzcor}
shows $C(t)$ in the ordered leaflet region close to
bilayer-bilayer interface ($z=0.3\;\textrm{nm}$) along with
the fit with two stretched/compressed exponentials. Assigning an
average decay time $\tau_{av,i} = \frac{\tau_i}{\beta_i} \Gamma \left(\frac{1}{\beta_i} \right)$, the fit gives a fast decay time of $0.03\;\textrm{ps}$ and
a slow decay time of $7.9\;\textrm{ns}$. At each $z$, we ensure that
the simulations are longer than the slow decay time.

\begin{figure}[hpbt]
\begin{center}
\includegraphics[width=\linewidth]{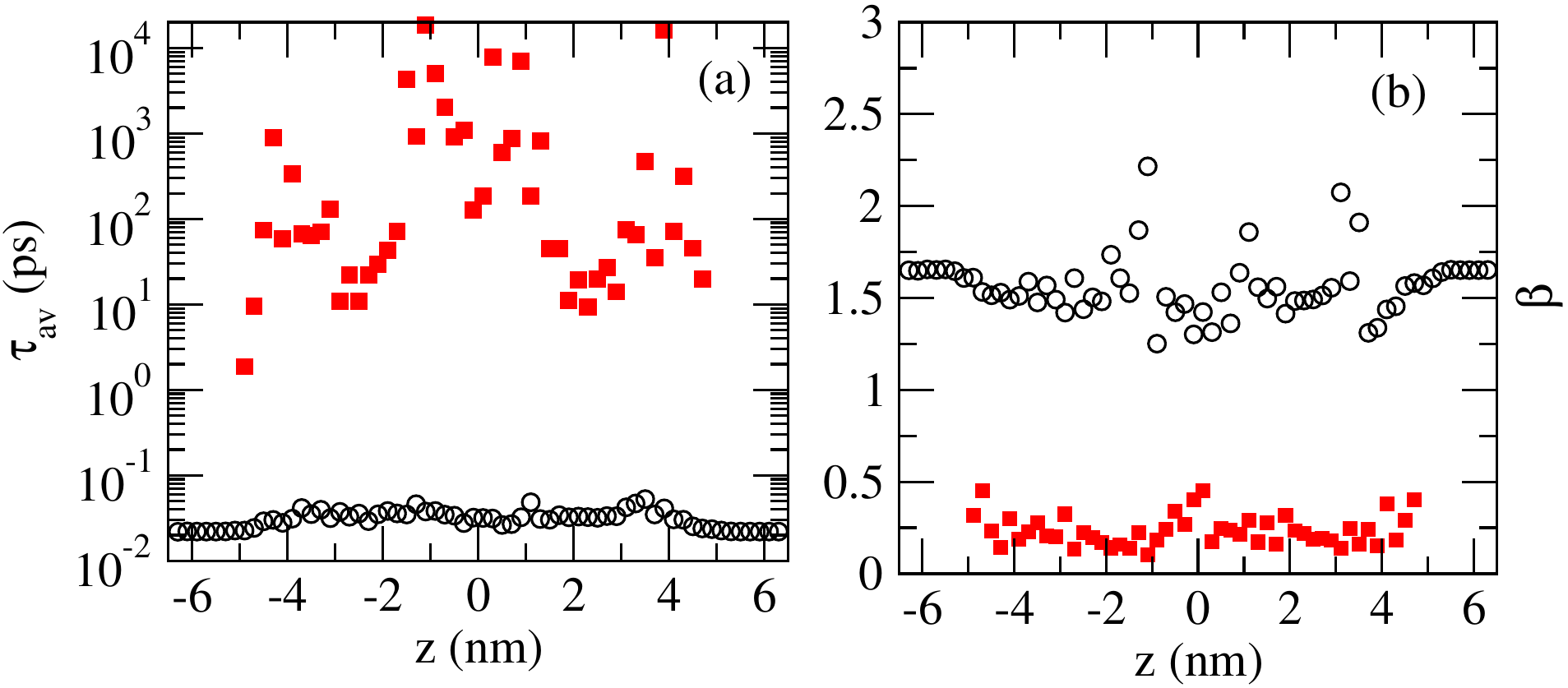}
\end{center}
\caption{Average decay times of $C(z,t)$ and exponents from  fitting of 
autocorrelation of $F_z (z,t)$,  at different distances $z$ from the double bilayer
center.}
\label{fig.fztime}
\end{figure}

Fig.~\ref{fig.fztime} shows the variation of the average decay times and 
the exponents with $z$. There is a fast decay in $C(z,t)$ at all $z$ 
with $\tau_{av} \simeq 0.03\;\textrm{ps}$ and exponent $\beta \simeq 1.5$
(open circles in Fig.~\ref{fig.fztime}). Inside the lipid double bilayer,
there is an additional slowly decaying component (filled squares in
Fig.~\ref{fig.fztime}) that follows a stretched exponential 
with $\tau_{av} \sim 20\;\textrm{ns}$ in the most ordered part of the
leaflets, with a stretching  exponent $\beta\simeq 0.2$.

Inside the ordered lipid leaflets, the $x-y$ diffusion of the
constrained water molecule is limited to $40\;\textrm{ns}$
timescale (the longest time simulated for the constrained water
simulations). We use six separate simulations with a different randomly 
chosen water molecules to calculate $\langle F_z(z)\rangle$ at a fixed $z$. Furthermore,
we use the antisymmetric property of $\langle F_z (z)\rangle$ to get 12 independent
estimates of $\langle F_z (z)\rangle$ at a given $z$. The error-bars in the
excess chemical potential are calculated from error of mean from these
12 estimates of $\langle F_z (z)\rangle$ at each $z$.


\begin{thebibliography}{43}%
\makeatletter
\providecommand \@ifxundefined [1]{%
 \@ifx{#1\undefined}
}%
\providecommand \@ifnum [1]{%
 \ifnum #1\expandafter \@firstoftwo
 \else \expandafter \@secondoftwo
 \fi
}%
\providecommand \@ifx [1]{%
 \ifx #1\expandafter \@firstoftwo
 \else \expandafter \@secondoftwo
 \fi
}%
\providecommand \natexlab [1]{#1}%
\providecommand \enquote  [1]{``#1''}%
\providecommand \bibnamefont  [1]{#1}%
\providecommand \bibfnamefont [1]{#1}%
\providecommand \citenamefont [1]{#1}%
\providecommand \href@noop [0]{\@secondoftwo}%
\providecommand \href [0]{\begingroup \@sanitize@url \@href}%
\providecommand \@href[1]{\@@startlink{#1}\@@href}%
\providecommand \@@href[1]{\endgroup#1\@@endlink}%
\providecommand \@sanitize@url [0]{\catcode `\\12\catcode `\$12\catcode
  `\&12\catcode `\#12\catcode `\^12\catcode `\_12\catcode `\%12\relax}%
\providecommand \@@startlink[1]{}%
\providecommand \@@endlink[0]{}%
\providecommand \url  [0]{\begingroup\@sanitize@url \@url }%
\providecommand \@url [1]{\endgroup\@href {#1}{\urlprefix }}%
\providecommand \urlprefix  [0]{URL }%
\providecommand \Eprint [0]{\href }%
\providecommand \doibase [0]{http://dx.doi.org/}%
\providecommand \selectlanguage [0]{\@gobble}%
\providecommand \bibinfo  [0]{\@secondoftwo}%
\providecommand \bibfield  [0]{\@secondoftwo}%
\providecommand \translation [1]{[#1]}%
\providecommand \BibitemOpen [0]{}%
\providecommand \bibitemStop [0]{}%
\providecommand \bibitemNoStop [0]{.\EOS\space}%
\providecommand \EOS [0]{\spacefactor3000\relax}%
\providecommand \BibitemShut  [1]{\csname bibitem#1\endcsname}%
\let\auto@bib@innerbib\@empty
\bibitem [{\citenamefont {Freinkel}\ and\ \citenamefont
  {Woodley}(2001)}]{freinkel.skin.01}%
  \BibitemOpen
  \bibinfo {editor} {\bibfnamefont {R.~K.}\ \bibnamefont {Freinkel}}\ and\
  \bibinfo {editor} {\bibfnamefont {D.~T.}\ \bibnamefont {Woodley}},\ eds.,\
  \href@noop {} {\emph {\bibinfo {title} {The {B}iology of the {S}kin}}}\
  (\bibinfo  {publisher} {Parthenon Publishing},\ \bibinfo {address} {London},\
  \bibinfo {year} {2001})\BibitemShut {NoStop}%
\bibitem [{\citenamefont {Elias}(2005)}]{elias.sc.rev.05}%
  \BibitemOpen
  \bibfield  {author} {\bibinfo {author} {\bibfnamefont {P.~M.}\ \bibnamefont
  {Elias}},\ }\href@noop {} {\bibfield  {journal} {\bibinfo  {journal} {J.
  Invest. Dermatol.}\ }\textbf {\bibinfo {volume} {125}},\ \bibinfo {pages}
  {183} (\bibinfo {year} {2005})}\BibitemShut {NoStop}%
\bibitem [{\citenamefont {Michaels}\ \emph {et~al.}(1975)\citenamefont
  {Michaels}, \citenamefont {Chandrasekaran},\ and\ \citenamefont
  {Shaw}}]{michaels.sc.brick.75}%
  \BibitemOpen
  \bibfield  {author} {\bibinfo {author} {\bibfnamefont {A.~S.}\ \bibnamefont
  {Michaels}}, \bibinfo {author} {\bibfnamefont {S.~K.}\ \bibnamefont
  {Chandrasekaran}}, \ and\ \bibinfo {author} {\bibfnamefont {J.~E.}\
  \bibnamefont {Shaw}},\ }\href@noop {} {\bibfield  {journal} {\bibinfo
  {journal} {AICHE J.}\ }\textbf {\bibinfo {volume} {21}},\ \bibinfo {pages}
  {985} (\bibinfo {year} {1975})}\BibitemShut {NoStop}%
\bibitem [{\citenamefont {{\AA}berg}\ \emph {et~al.}(2008)\citenamefont
  {{\AA}berg}, \citenamefont {Wennerstr{\"o}m},\ and\ \citenamefont
  {Sparr}}]{aberg.08}%
  \BibitemOpen
  \bibfield  {author} {\bibinfo {author} {\bibfnamefont {C.}~\bibnamefont
  {{\AA}berg}}, \bibinfo {author} {\bibfnamefont {H.}~\bibnamefont
  {Wennerstr{\"o}m}}, \ and\ \bibinfo {author} {\bibfnamefont {E.}~\bibnamefont
  {Sparr}},\ }\href@noop {} {\bibfield  {journal} {\bibinfo  {journal}
  {Langmuir}\ }\textbf {\bibinfo {volume} {84}},\ \bibinfo {pages} {8061}
  (\bibinfo {year} {2008})}\BibitemShut {NoStop}%
\bibitem [{\citenamefont {Moghimi}\ \emph {et~al.}(1996)\citenamefont
  {Moghimi}, \citenamefont {Williams},\ and\ \citenamefont
  {Barry}}]{moghimi.jpharm.96}%
  \BibitemOpen
  \bibfield  {author} {\bibinfo {author} {\bibfnamefont {H.~R.}\ \bibnamefont
  {Moghimi}}, \bibinfo {author} {\bibfnamefont {A.~C.}\ \bibnamefont
  {Williams}}, \ and\ \bibinfo {author} {\bibfnamefont {B.~W.}\ \bibnamefont
  {Barry}},\ }\href@noop {} {\bibfield  {journal} {\bibinfo  {journal} {Int. J.
  Pharm.}\ }\textbf {\bibinfo {volume} {131}},\ \bibinfo {pages} {117}
  (\bibinfo {year} {1996})}\BibitemShut {NoStop}%
\bibitem [{\citenamefont {Sparr}\ and\ \citenamefont
  {Wennerstr{\"o}m}(2000)}]{sparr.col.00}%
  \BibitemOpen
  \bibfield  {author} {\bibinfo {author} {\bibfnamefont {E.}~\bibnamefont
  {Sparr}}\ and\ \bibinfo {author} {\bibfnamefont {H.}~\bibnamefont
  {Wennerstr{\"o}m}},\ }\href@noop {} {\bibfield  {journal} {\bibinfo
  {journal} {Col. Surf. B: Biointerfaces}\ }\textbf {\bibinfo {volume} {19}},\
  \bibinfo {pages} {103} (\bibinfo {year} {2000})}\BibitemShut {NoStop}%
\bibitem [{\citenamefont {Bouwstra}\ \emph {et~al.}(1991)\citenamefont
  {Bouwstra}, \citenamefont {Gooris}, \citenamefont {ver~der Spek},\ and\
  \citenamefont {Bras}}]{bouwstra.jid.91}%
  \BibitemOpen
  \bibfield  {author} {\bibinfo {author} {\bibfnamefont {J.~A.}\ \bibnamefont
  {Bouwstra}}, \bibinfo {author} {\bibfnamefont {G.~S.}\ \bibnamefont
  {Gooris}}, \bibinfo {author} {\bibfnamefont {J.~A.}\ \bibnamefont {ver~der
  Spek}}, \ and\ \bibinfo {author} {\bibfnamefont {W.}~\bibnamefont {Bras}},\
  }\href@noop {} {\bibfield  {journal} {\bibinfo  {journal} {J. Invest. Derm.}\
  }\textbf {\bibinfo {volume} {97}},\ \bibinfo {pages} {1005} (\bibinfo {year}
  {1991})}\BibitemShut {NoStop}%
\bibitem [{\citenamefont {Mak}\ \emph {et~al.}(1991)\citenamefont {Mak},
  \citenamefont {Potts},\ and\ \citenamefont {Guy}}]{mak.pharmres.91}%
  \BibitemOpen
  \bibfield  {author} {\bibinfo {author} {\bibfnamefont {V.~H.~W.}\
  \bibnamefont {Mak}}, \bibinfo {author} {\bibfnamefont {R.~O.}\ \bibnamefont
  {Potts}}, \ and\ \bibinfo {author} {\bibfnamefont {R.~H.}\ \bibnamefont
  {Guy}},\ }\href@noop {} {\bibfield  {journal} {\bibinfo  {journal} {Pharm.
  Res.}\ }\textbf {\bibinfo {volume} {8}},\ \bibinfo {pages} {1064} (\bibinfo
  {year} {1991})}\BibitemShut {NoStop}%
\bibitem [{\citenamefont {Kiselev}\ \emph {et~al.}(2005)\citenamefont
  {Kiselev}, \citenamefont {Ryabova}, \citenamefont {Balagurov}, \citenamefont
  {Dante}, \citenamefont {Hauss}, \citenamefont {Zbytovska}, \citenamefont
  {Wartewig},\ and\ \citenamefont {Neubert}}]{kiselev.EBJ.05}%
  \BibitemOpen
  \bibfield  {author} {\bibinfo {author} {\bibfnamefont {M.~A.}\ \bibnamefont
  {Kiselev}}, \bibinfo {author} {\bibfnamefont {N.~Y.}\ \bibnamefont
  {Ryabova}}, \bibinfo {author} {\bibfnamefont {A.~M.}\ \bibnamefont
  {Balagurov}}, \bibinfo {author} {\bibfnamefont {S.}~\bibnamefont {Dante}},
  \bibinfo {author} {\bibfnamefont {T.}~\bibnamefont {Hauss}}, \bibinfo
  {author} {\bibfnamefont {J.}~\bibnamefont {Zbytovska}}, \bibinfo {author}
  {\bibfnamefont {S.}~\bibnamefont {Wartewig}}, \ and\ \bibinfo {author}
  {\bibfnamefont {R.~H.~H.}\ \bibnamefont {Neubert}},\ }\href@noop {}
  {\bibfield  {journal} {\bibinfo  {journal} {Eur. Biophys. J.}\ }\textbf
  {\bibinfo {volume} {34}},\ \bibinfo {pages} {1030} (\bibinfo {year}
  {2005})}\BibitemShut {NoStop}%
\bibitem [{\citenamefont {Swartzendruber}\ \emph {et~al.}(1989)\citenamefont
  {Swartzendruber}, \citenamefont {Wertz}, \citenamefont {Kitko}, \citenamefont
  {Madison},\ and\ \citenamefont {Downing}}]{swartzendruber.jid.89}%
  \BibitemOpen
  \bibfield  {author} {\bibinfo {author} {\bibfnamefont {D.~C.}\ \bibnamefont
  {Swartzendruber}}, \bibinfo {author} {\bibfnamefont {P.~W.}\ \bibnamefont
  {Wertz}}, \bibinfo {author} {\bibfnamefont {D.~J.}\ \bibnamefont {Kitko}},
  \bibinfo {author} {\bibfnamefont {K.~C.}\ \bibnamefont {Madison}}, \ and\
  \bibinfo {author} {\bibfnamefont {D.~T.}\ \bibnamefont {Downing}},\
  }\href@noop {} {\bibfield  {journal} {\bibinfo  {journal} {J. Invest. Derm.}\
  }\textbf {\bibinfo {volume} {92}},\ \bibinfo {pages} {251} (\bibinfo {year}
  {1989})}\BibitemShut {NoStop}%
\bibitem [{\citenamefont {Forslind}(1994)}]{forslind.adv.94}%
  \BibitemOpen
  \bibfield  {author} {\bibinfo {author} {\bibfnamefont {B.~O.}\ \bibnamefont
  {Forslind}},\ }\href@noop {} {\bibfield  {journal} {\bibinfo  {journal} {Acta
  Dermato Venerol}\ }\textbf {\bibinfo {volume} {74}},\ \bibinfo {pages} {1}
  (\bibinfo {year} {1994})}\BibitemShut {NoStop}%
\bibitem [{\citenamefont {Bouwstra}\ \emph {et~al.}(1998)\citenamefont
  {Bouwstra}, \citenamefont {Gooris}, \citenamefont {Dubbelaar}, \citenamefont
  {Weerheim}, \citenamefont {Ijzerman},\ and\ \citenamefont
  {Ponec}}]{bouwstra.jlr.98}%
  \BibitemOpen
  \bibfield  {author} {\bibinfo {author} {\bibfnamefont {J.~A.}\ \bibnamefont
  {Bouwstra}}, \bibinfo {author} {\bibfnamefont {G.~S.}\ \bibnamefont
  {Gooris}}, \bibinfo {author} {\bibfnamefont {F.~E.~R.}\ \bibnamefont
  {Dubbelaar}}, \bibinfo {author} {\bibfnamefont {A.~M.}\ \bibnamefont
  {Weerheim}}, \bibinfo {author} {\bibfnamefont {A.~P.}\ \bibnamefont
  {Ijzerman}}, \ and\ \bibinfo {author} {\bibfnamefont {M.}~\bibnamefont
  {Ponec}},\ }\href@noop {} {\bibfield  {journal} {\bibinfo  {journal} {J.
  Lipid Res.}\ }\textbf {\bibinfo {volume} {39}},\ \bibinfo {pages} {186}
  (\bibinfo {year} {1998})}\BibitemShut {NoStop}%
\bibitem [{\citenamefont {McIntosh}(2003)}]{mcintosh.bpj.03}%
  \BibitemOpen
  \bibfield  {author} {\bibinfo {author} {\bibfnamefont {T.~J.}\ \bibnamefont
  {McIntosh}},\ }\href@noop {} {\bibfield  {journal} {\bibinfo  {journal}
  {Biophys. J.}\ }\textbf {\bibinfo {volume} {85}},\ \bibinfo {pages} {1675}
  (\bibinfo {year} {2003})}\BibitemShut {NoStop}%
\bibitem [{\citenamefont {Hill}\ and\ \citenamefont
  {Wertz}(2003)}]{hill.bba.03}%
  \BibitemOpen
  \bibfield  {author} {\bibinfo {author} {\bibfnamefont {J.~R.}\ \bibnamefont
  {Hill}}\ and\ \bibinfo {author} {\bibfnamefont {P.~W.}\ \bibnamefont
  {Wertz}},\ }\href@noop {} {\bibfield  {journal} {\bibinfo  {journal}
  {Biochim. Biophys. Acta - Biomembranes}\ }\textbf {\bibinfo {volume}
  {1616}},\ \bibinfo {pages} {121} (\bibinfo {year} {2003})}\BibitemShut
  {NoStop}%
\bibitem [{\citenamefont {Schr\:oter}\ \emph {et~al.}(2009)\citenamefont
  {Schr\:oter}, \citenamefont {Kessner}, \citenamefont {Kiselev}, \citenamefont
  {Hau\ss}, \citenamefont {Dante},\ and\ \citenamefont
  {Neubert}}]{schroter.bpj.09}%
  \BibitemOpen
  \bibfield  {author} {\bibinfo {author} {\bibfnamefont {A.}~\bibnamefont
  {Schr\:oter}}, \bibinfo {author} {\bibfnamefont {D.}~\bibnamefont {Kessner}},
  \bibinfo {author} {\bibfnamefont {M.}~\bibnamefont {Kiselev}}, \bibinfo
  {author} {\bibfnamefont {T.}~\bibnamefont {Hau\ss}}, \bibinfo {author}
  {\bibfnamefont {S.}~\bibnamefont {Dante}}, \ and\ \bibinfo {author}
  {\bibfnamefont {R.~H.}\ \bibnamefont {Neubert}},\ }\href@noop {} {\bibfield
  {journal} {\bibinfo  {journal} {Biophys. J.}\ }\textbf {\bibinfo {volume}
  {97}},\ \bibinfo {pages} {1104} (\bibinfo {year} {2009})}\BibitemShut
  {NoStop}%
\bibitem [{\citenamefont {Iwai}\ \emph {et~al.}(2012)\citenamefont {Iwai},
  \citenamefont {Han}, \citenamefont {den Hollander}, \citenamefont {Svensson},
  \citenamefont {\:Ofverstedt}, \citenamefont {Anwar}, \citenamefont {Brewer},
  \citenamefont {Bloksgaard}, \citenamefont {Laloeuf}, \citenamefont {Nosek},
  \citenamefont {Masich}, \citenamefont {Bagatolli}, \citenamefont {Skoglund},\
  and\ \citenamefont {L.}}]{iwai.jid.12}%
  \BibitemOpen
  \bibfield  {author} {\bibinfo {author} {\bibfnamefont {I.}~\bibnamefont
  {Iwai}}, \bibinfo {author} {\bibfnamefont {H.}~\bibnamefont {Han}}, \bibinfo
  {author} {\bibfnamefont {L.}~\bibnamefont {den Hollander}}, \bibinfo {author}
  {\bibfnamefont {S.}~\bibnamefont {Svensson}}, \bibinfo {author}
  {\bibfnamefont {L.}~\bibnamefont {\:Ofverstedt}}, \bibinfo {author}
  {\bibfnamefont {J.}~\bibnamefont {Anwar}}, \bibinfo {author} {\bibfnamefont
  {J.}~\bibnamefont {Brewer}}, \bibinfo {author} {\bibfnamefont
  {M.}~\bibnamefont {Bloksgaard}}, \bibinfo {author} {\bibfnamefont
  {A.}~\bibnamefont {Laloeuf}}, \bibinfo {author} {\bibfnamefont
  {D.}~\bibnamefont {Nosek}}, \bibinfo {author} {\bibfnamefont
  {S.}~\bibnamefont {Masich}}, \bibinfo {author} {\bibfnamefont
  {L.}~\bibnamefont {Bagatolli}}, \bibinfo {author} {\bibfnamefont
  {U.}~\bibnamefont {Skoglund}}, \ and\ \bibinfo {author} {\bibfnamefont
  {N.}~\bibnamefont {L.}},\ }\href@noop {} {\bibfield  {journal} {\bibinfo
  {journal} {J. Invest. Derm.}\ }\textbf {\bibinfo {volume} {132}},\ \bibinfo
  {pages} {2215} (\bibinfo {year} {2012})}\BibitemShut {NoStop}%
\bibitem [{\citenamefont {Bouwstra}\ \emph {et~al.}(2003)\citenamefont
  {Bouwstra}, \citenamefont {de~Graaff}, \citenamefont {Gooris}, \citenamefont
  {Nijsse}, \citenamefont {Wiechers},\ and\ \citenamefont {van Aelst
  Adriaan~C.}}]{bouwstra.jid.03}%
  \BibitemOpen
  \bibfield  {author} {\bibinfo {author} {\bibfnamefont {J.~A.}\ \bibnamefont
  {Bouwstra}}, \bibinfo {author} {\bibfnamefont {A.}~\bibnamefont {de~Graaff}},
  \bibinfo {author} {\bibfnamefont {G.~S.}\ \bibnamefont {Gooris}}, \bibinfo
  {author} {\bibfnamefont {J.}~\bibnamefont {Nijsse}}, \bibinfo {author}
  {\bibfnamefont {J.~W.}\ \bibnamefont {Wiechers}}, \ and\ \bibinfo {author}
  {\bibnamefont {van Aelst Adriaan~C.}},\ }\href@noop {} {\bibfield  {journal}
  {\bibinfo  {journal} {J. Invest. Derm.}\ }\textbf {\bibinfo {volume} {120}},\
  \bibinfo {pages} {750} (\bibinfo {year} {2003})}\BibitemShut {NoStop}%
\bibitem [{\citenamefont {Richter}\ \emph {et~al.}(2001)\citenamefont
  {Richter}, \citenamefont {M\"uller}, \citenamefont {Schwarz}, \citenamefont
  {Wepf},\ and\ \citenamefont {Wiesendanger}}]{richter.appphysa.01}%
  \BibitemOpen
  \bibfield  {author} {\bibinfo {author} {\bibfnamefont {T.}~\bibnamefont
  {Richter}}, \bibinfo {author} {\bibfnamefont {J.~H.}\ \bibnamefont
  {M\"uller}}, \bibinfo {author} {\bibfnamefont {U.~D.}\ \bibnamefont
  {Schwarz}}, \bibinfo {author} {\bibfnamefont {R.}~\bibnamefont {Wepf}}, \
  and\ \bibinfo {author} {\bibfnamefont {R.}~\bibnamefont {Wiesendanger}},\
  }\href@noop {} {\bibfield  {journal} {\bibinfo  {journal} {Appl. Phys. A}\
  }\textbf {\bibinfo {volume} {72}},\ \bibinfo {pages} {S125} (\bibinfo {year}
  {2001})}\BibitemShut {NoStop}%
\bibitem [{\citenamefont {H\"oltje}\ \emph {et~al.}(2001)\citenamefont
  {H\"oltje}, \citenamefont {F\"orster}, \citenamefont {Brandt}, \citenamefont
  {Engels}, \citenamefont {von Rybinski},\ and\ \citenamefont
  {H\"oltje}}]{holtje.fachol.01}%
  \BibitemOpen
  \bibfield  {author} {\bibinfo {author} {\bibfnamefont {M.}~\bibnamefont
  {H\"oltje}}, \bibinfo {author} {\bibfnamefont {T.}~\bibnamefont {F\"orster}},
  \bibinfo {author} {\bibfnamefont {B.}~\bibnamefont {Brandt}}, \bibinfo
  {author} {\bibfnamefont {T.}~\bibnamefont {Engels}}, \bibinfo {author}
  {\bibfnamefont {W.}~\bibnamefont {von Rybinski}}, \ and\ \bibinfo {author}
  {\bibfnamefont {H.-D.}\ \bibnamefont {H\"oltje}},\ }\href@noop {} {\bibfield
  {journal} {\bibinfo  {journal} {Biochim. Biophys. Acta}\ }\textbf {\bibinfo
  {volume} {1511}},\ \bibinfo {pages} {156} (\bibinfo {year}
  {2001})}\BibitemShut {NoStop}%
\bibitem [{\citenamefont {Pandit}\ and\ \citenamefont
  {Scott}(2006)}]{pandit.cer2.06}%
  \BibitemOpen
  \bibfield  {author} {\bibinfo {author} {\bibfnamefont {S.~A.}\ \bibnamefont
  {Pandit}}\ and\ \bibinfo {author} {\bibfnamefont {H.~L.}\ \bibnamefont
  {Scott}},\ }\href@noop {} {\bibfield  {journal} {\bibinfo  {journal} {J.
  Chem. Phys.}\ }\textbf {\bibinfo {volume} {124}},\ \bibinfo {pages} {014708}
  (\bibinfo {year} {2006})}\BibitemShut {NoStop}%
\bibitem [{\citenamefont {Notman}\ \emph {et~al.}(2007)\citenamefont {Notman},
  \citenamefont {den Otter}, \citenamefont {Noro}, \citenamefont {Briels},\
  and\ \citenamefont {Anwar}}]{notman.dmso.07}%
  \BibitemOpen
  \bibfield  {author} {\bibinfo {author} {\bibfnamefont {R.}~\bibnamefont
  {Notman}}, \bibinfo {author} {\bibfnamefont {W.~K.}\ \bibnamefont {den
  Otter}}, \bibinfo {author} {\bibfnamefont {M.~G.}\ \bibnamefont {Noro}},
  \bibinfo {author} {\bibfnamefont {W.~J.}\ \bibnamefont {Briels}}, \ and\
  \bibinfo {author} {\bibfnamefont {J.}~\bibnamefont {Anwar}},\ }\href@noop {}
  {\bibfield  {journal} {\bibinfo  {journal} {Biophys. J.}\ }\textbf {\bibinfo
  {volume} {93}},\ \bibinfo {pages} {2056} (\bibinfo {year}
  {2007})}\BibitemShut {NoStop}%
\bibitem [{\citenamefont {Das}\ \emph {et~al.}(2009)\citenamefont {Das},
  \citenamefont {Noro},\ and\ \citenamefont {Olmsted}}]{das.bpj.09}%
  \BibitemOpen
  \bibfield  {author} {\bibinfo {author} {\bibfnamefont {C.}~\bibnamefont
  {Das}}, \bibinfo {author} {\bibfnamefont {M.~G.}\ \bibnamefont {Noro}}, \
  and\ \bibinfo {author} {\bibfnamefont {P.~D.}\ \bibnamefont {Olmsted}},\
  }\href@noop {} {\bibfield  {journal} {\bibinfo  {journal} {Biophys. J.}\
  }\textbf {\bibinfo {volume} {97}},\ \bibinfo {pages} {1941} (\bibinfo {year}
  {2009})}\BibitemShut {NoStop}%
\bibitem [{\citenamefont {Das}\ \emph {et~al.}(2010)\citenamefont {Das},
  \citenamefont {Olmsted},\ and\ \citenamefont {Noro}}]{das.smat.10}%
  \BibitemOpen
  \bibfield  {author} {\bibinfo {author} {\bibfnamefont {C.}~\bibnamefont
  {Das}}, \bibinfo {author} {\bibfnamefont {P.~D.}\ \bibnamefont {Olmsted}}, \
  and\ \bibinfo {author} {\bibfnamefont {M.~G.}\ \bibnamefont {Noro}},\
  }\href@noop {} {\bibfield  {journal} {\bibinfo  {journal} {Soft Matter}\
  }\textbf {\bibinfo {volume} {5}},\ \bibinfo {pages} {4549} (\bibinfo {year}
  {2010})}\BibitemShut {NoStop}%
\bibitem [{\citenamefont {Das}\ \emph {et~al.}(2013)\citenamefont {Das},
  \citenamefont {Noro},\ and\ \citenamefont {Olmsted}}]{das2013lamellar}%
  \BibitemOpen
  \bibfield  {author} {\bibinfo {author} {\bibfnamefont {C.}~\bibnamefont
  {Das}}, \bibinfo {author} {\bibfnamefont {M.~G.}\ \bibnamefont {Noro}}, \
  and\ \bibinfo {author} {\bibfnamefont {P.~D.}\ \bibnamefont {Olmsted}},\
  }\href@noop {} {\bibfield  {journal} {\bibinfo  {journal} {Phys. Rev. Lett.}\
  }\textbf {\bibinfo {volume} {111}},\ \bibinfo {pages} {148101} (\bibinfo
  {year} {2013})}\BibitemShut {NoStop}%
\bibitem [{\citenamefont {Engelbrecht}\ \emph {et~al.}(2011)\citenamefont
  {Engelbrecht}, \citenamefont {Hau\ss}, \citenamefont {S\"u\ss}, \citenamefont
  {Vogel}, \citenamefont {Roark}, \citenamefont {Feller}, \citenamefont
  {Neubert},\ and\ \citenamefont {Dobner}}]{engelbrecht.smat.11}%
  \BibitemOpen
  \bibfield  {author} {\bibinfo {author} {\bibfnamefont {T.}~\bibnamefont
  {Engelbrecht}}, \bibinfo {author} {\bibfnamefont {T.}~\bibnamefont {Hau\ss}},
  \bibinfo {author} {\bibfnamefont {K.}~\bibnamefont {S\"u\ss}}, \bibinfo
  {author} {\bibfnamefont {A.}~\bibnamefont {Vogel}}, \bibinfo {author}
  {\bibfnamefont {M.}~\bibnamefont {Roark}}, \bibinfo {author} {\bibfnamefont
  {S.~E.}\ \bibnamefont {Feller}}, \bibinfo {author} {\bibfnamefont {R.~H.~H.}\
  \bibnamefont {Neubert}}, \ and\ \bibinfo {author} {\bibfnamefont
  {B.}~\bibnamefont {Dobner}},\ }\href@noop {} {\bibfield  {journal} {\bibinfo
  {journal} {Soft Matter}\ }\textbf {\bibinfo {volume} {7}},\ \bibinfo {pages}
  {8998} (\bibinfo {year} {2011})}\BibitemShut {NoStop}%
\bibitem [{\citenamefont {Ryckaert}\ and\ \citenamefont
  {Bellemans}(1975)}]{ryckaert.ff.75}%
  \BibitemOpen
  \bibfield  {author} {\bibinfo {author} {\bibfnamefont {J.-P.}\ \bibnamefont
  {Ryckaert}}\ and\ \bibinfo {author} {\bibfnamefont {A.}~\bibnamefont
  {Bellemans}},\ }\href@noop {} {\bibfield  {journal} {\bibinfo  {journal}
  {Chem. Phys. Lett.}\ }\textbf {\bibinfo {volume} {30}},\ \bibinfo {pages}
  {123} (\bibinfo {year} {1975})}\BibitemShut {NoStop}%
\bibitem [{\citenamefont {Jorgensen}\ and\ \citenamefont
  {Tirado-Rives}(1988)}]{jorgensen.ff.88}%
  \BibitemOpen
  \bibfield  {author} {\bibinfo {author} {\bibfnamefont {W.}~\bibnamefont
  {Jorgensen}}\ and\ \bibinfo {author} {\bibfnamefont {J.}~\bibnamefont
  {Tirado-Rives}},\ }\href@noop {} {\bibfield  {journal} {\bibinfo  {journal}
  {J. Am. Chem. Soc.}\ }\textbf {\bibinfo {volume} {110}},\ \bibinfo {pages}
  {1657} (\bibinfo {year} {1988})}\BibitemShut {NoStop}%
\bibitem [{\citenamefont {Chiu}\ \emph {et~al.}(1995)\citenamefont {Chiu},
  \citenamefont {Clark}, \citenamefont {Balaji}, \citenamefont {Subramaniam},
  \citenamefont {Scott},\ and\ \citenamefont {Jackobsson}}]{chiu.ff.95}%
  \BibitemOpen
  \bibfield  {author} {\bibinfo {author} {\bibfnamefont {S.~W.}\ \bibnamefont
  {Chiu}}, \bibinfo {author} {\bibfnamefont {M.}~\bibnamefont {Clark}},
  \bibinfo {author} {\bibfnamefont {V.}~\bibnamefont {Balaji}}, \bibinfo
  {author} {\bibfnamefont {S.}~\bibnamefont {Subramaniam}}, \bibinfo {author}
  {\bibfnamefont {H.~L.}\ \bibnamefont {Scott}}, \ and\ \bibinfo {author}
  {\bibfnamefont {E.}~\bibnamefont {Jackobsson}},\ }\href@noop {} {\bibfield
  {journal} {\bibinfo  {journal} {Biophys. J.}\ }\textbf {\bibinfo {volume}
  {69}},\ \bibinfo {pages} {1230} (\bibinfo {year} {1995})}\BibitemShut
  {NoStop}%
\bibitem [{\citenamefont {Berger}\ \emph {et~al.}(1997)\citenamefont {Berger},
  \citenamefont {Edholm},\ and\ \citenamefont {J\"ahnig}}]{berger.ff.97}%
  \BibitemOpen
  \bibfield  {author} {\bibinfo {author} {\bibfnamefont {O.}~\bibnamefont
  {Berger}}, \bibinfo {author} {\bibfnamefont {O.}~\bibnamefont {Edholm}}, \
  and\ \bibinfo {author} {\bibfnamefont {F.}~\bibnamefont {J\"ahnig}},\
  }\href@noop {} {\bibfield  {journal} {\bibinfo  {journal} {Biophys. J.}\
  }\textbf {\bibinfo {volume} {72}},\ \bibinfo {pages} {2002} (\bibinfo {year}
  {1997})}\BibitemShut {NoStop}%
\bibitem [{\citenamefont {Berendsen}\ \emph {et~al.}(1981)\citenamefont
  {Berendsen}, \citenamefont {Postma}, \citenamefont {van Gunsteren},\ and\
  \citenamefont {Hermans}}]{spcwater}%
  \BibitemOpen
  \bibfield  {author} {\bibinfo {author} {\bibfnamefont {H.~J.~C.}\
  \bibnamefont {Berendsen}}, \bibinfo {author} {\bibfnamefont {J.~P.~M.}\
  \bibnamefont {Postma}}, \bibinfo {author} {\bibfnamefont {W.~F.}\
  \bibnamefont {van Gunsteren}}, \ and\ \bibinfo {author} {\bibfnamefont
  {J.}~\bibnamefont {Hermans}},\ }in\ \href@noop {} {\emph {\bibinfo
  {booktitle} {Intermolecular Forces}}},\ \bibinfo {editor} {edited by\
  \bibinfo {editor} {\bibfnamefont {B.}~\bibnamefont {Pullman}}}\ (\bibinfo
  {publisher} {Reidel},\ \bibinfo {address} {Dordrecht},\ \bibinfo {year}
  {1981})\ pp.\ \bibinfo {pages} {331--342}\BibitemShut {NoStop}%
\bibitem [{Sup()}]{Supplementary}%
  \BibitemOpen
  \href@noop {} {}\bibinfo {note} {Supplementary Material at
  \url{http://link.aps.org/supplemental/xxx} contains an animation, and details
  about calculating hydrogen bonds, flip-flop times, diffusion, and the excess
  chemical potential of water. Other animations also available on
  \url{http://goo.gl/qqMzrE}.}\BibitemShut {Stop}%
\bibitem [{\citenamefont {Berendsen}\ \emph {et~al.}(1995)\citenamefont
  {Berendsen}, \citenamefont {van~der Spoel},\ and\ \citenamefont {van
  Drunen}}]{gromacs95}%
  \BibitemOpen
  \bibfield  {author} {\bibinfo {author} {\bibfnamefont {H.~J.~C.}\
  \bibnamefont {Berendsen}}, \bibinfo {author} {\bibfnamefont {D.}~\bibnamefont
  {van~der Spoel}}, \ and\ \bibinfo {author} {\bibfnamefont {R.}~\bibnamefont
  {van Drunen}},\ }\href@noop {} {\bibfield  {journal} {\bibinfo  {journal}
  {Comp. Phys. Comm.}\ }\textbf {\bibinfo {volume} {91}},\ \bibinfo {pages}
  {43} (\bibinfo {year} {1995})}\BibitemShut {NoStop}%
\bibitem [{\citenamefont {van~der Spoel}\ \emph {et~al.}(2005)\citenamefont
  {van~der Spoel}, \citenamefont {Lindahl}, \citenamefont {Hess}, \citenamefont
  {Groenhof}, \citenamefont {Mark},\ and\ \citenamefont
  {Berendsen}}]{gromacs05}%
  \BibitemOpen
  \bibfield  {author} {\bibinfo {author} {\bibfnamefont {D.}~\bibnamefont
  {van~der Spoel}}, \bibinfo {author} {\bibfnamefont {E.}~\bibnamefont
  {Lindahl}}, \bibinfo {author} {\bibfnamefont {B.}~\bibnamefont {Hess}},
  \bibinfo {author} {\bibfnamefont {G.}~\bibnamefont {Groenhof}}, \bibinfo
  {author} {\bibfnamefont {A.~E.}\ \bibnamefont {Mark}}, \ and\ \bibinfo
  {author} {\bibfnamefont {H.~J.~C.}\ \bibnamefont {Berendsen}},\ }\href@noop
  {} {\bibfield  {journal} {\bibinfo  {journal} {J. Comp. Chem.}\ }\textbf
  {\bibinfo {volume} {26}},\ \bibinfo {pages} {1701} (\bibinfo {year}
  {2005})}\BibitemShut {NoStop}%
\bibitem [{\citenamefont {Marrink}\ and\ \citenamefont
  {Berendsen}(1994)}]{marrink.jpc.94}%
  \BibitemOpen
  \bibfield  {author} {\bibinfo {author} {\bibfnamefont {S.~J.}\ \bibnamefont
  {Marrink}}\ and\ \bibinfo {author} {\bibfnamefont {H.~J.~C.}\ \bibnamefont
  {Berendsen}},\ }\href@noop {} {\bibfield  {journal} {\bibinfo  {journal} {J.
  Phys. Chem.}\ }\textbf {\bibinfo {volume} {98}},\ \bibinfo {pages} {4155}
  (\bibinfo {year} {1994})}\BibitemShut {NoStop}%
\bibitem [{Note1()}]{Note1}%
  \BibitemOpen
  \bibinfo {note} {In steady state, detailed balance implies that $N_{21}\equiv
  n_1k_{21}\protect \tmspace +\thinmuskip {.1667em}dt = N_{12}$, where $n_i$ is
  the number in a given zone, $k_{ji}$ is the rate, per molecule, that a
  molecule in zone $i$ transits to zone $j$, and $N_{ji}$ is the total number
  that transit in a time $dt$. Hence, if there are $N$ total transitions
  between two zones in time $T$, then the characteristic time $\tau
  _{1\rightarrow 2}=k_{21}^{-1} =2 n_1 T/N$. For $N=322$ flip events in $T=1\mu
  \protect \textrm {s}$, $n_{\protect \textrm {\relax \protect \fontsize
  {5}{6}\protect \selectfont O}} \simeq 102.7$ and $n_{\protect \textrm {\relax
  \protect \fontsize {5}{6}\protect \selectfont D}} \simeq 9.3$, giving $\tau
  _{\protect \textrm {\relax \protect \fontsize {5}{6}\protect \selectfont O}
  \rightarrow \protect \textrm {\relax \protect \fontsize {5}{6}\protect
  \selectfont D}} = (2 \times 102.7/322) \mu \protect \textrm {s} = 0.64 \mu
  \protect \textrm {s}$ and $\tau _{\protect \textrm {\relax \protect \fontsize
  {5}{6}\protect \selectfont D} \rightarrow \protect \textrm {\relax \protect
  \fontsize {5}{6}\protect \selectfont O}} =0.06 \mu \protect \textrm {s}$. For
  flip-flop events we consider all 112 CHOL molecules to be members of one of
  the two leaflets (exploring an ordered and a disordered zone) until the
  flip-flop event takes place (to another ordered zone). Hence, 6 flip-flop
  events in $1\mu \protect \textrm {s}$ leads to $\tau _{\protect \textit {ff}}
  = 2\times (112/2)/6 = 18.7 \mu \protect \textrm {s}$.}\BibitemShut {Stop}%
\bibitem [{\citenamefont {Garg}\ \emph {et~al.}(2011)\citenamefont {Garg},
  \citenamefont {Porcar}, \citenamefont {Woodka}, \citenamefont {Butler},\ and\
  \citenamefont {Perez-Salas}}]{Garg.BPJ.11}%
  \BibitemOpen
  \bibfield  {author} {\bibinfo {author} {\bibfnamefont {S.}~\bibnamefont
  {Garg}}, \bibinfo {author} {\bibfnamefont {L.}~\bibnamefont {Porcar}},
  \bibinfo {author} {\bibfnamefont {A.~C.}\ \bibnamefont {Woodka}}, \bibinfo
  {author} {\bibfnamefont {P.~D.}\ \bibnamefont {Butler}}, \ and\ \bibinfo
  {author} {\bibfnamefont {U.}~\bibnamefont {Perez-Salas}},\ }\href@noop {}
  {\bibfield  {journal} {\bibinfo  {journal} {Biophys. J.}\ }\textbf {\bibinfo
  {volume} {101}},\ \bibinfo {pages} {370 } (\bibinfo {year}
  {2011})}\BibitemShut {NoStop}%
\bibitem [{\citenamefont {Bruckner}\ \emph {et~al.}(2009)\citenamefont
  {Bruckner}, \citenamefont {Mansy}, \citenamefont {Ricardo}, \citenamefont
  {Mahadevan},\ and\ \citenamefont {Szostak}}]{Bruckner.BPJ.09}%
  \BibitemOpen
  \bibfield  {author} {\bibinfo {author} {\bibfnamefont {R.~J.}\ \bibnamefont
  {Bruckner}}, \bibinfo {author} {\bibfnamefont {S.~S.}\ \bibnamefont {Mansy}},
  \bibinfo {author} {\bibfnamefont {A.}~\bibnamefont {Ricardo}}, \bibinfo
  {author} {\bibfnamefont {L.}~\bibnamefont {Mahadevan}}, \ and\ \bibinfo
  {author} {\bibfnamefont {J.~W.}\ \bibnamefont {Szostak}},\ }\href@noop {}
  {\bibfield  {journal} {\bibinfo  {journal} {Biophys. J.}\ }\textbf {\bibinfo
  {volume} {97}},\ \bibinfo {pages} {3113} (\bibinfo {year}
  {2009})}\BibitemShut {NoStop}%
\bibitem [{\citenamefont {Drew~Bennett}\ \emph {et~al.}(2009)\citenamefont
  {Drew~Bennett}, \citenamefont {MacCallum}, \citenamefont {Hinner},
  \citenamefont {Marrink},\ and\ \citenamefont {Tieleman}}]{Bennet.jacs.09}%
  \BibitemOpen
  \bibfield  {author} {\bibinfo {author} {\bibfnamefont {W.~F.}\ \bibnamefont
  {Drew~Bennett}}, \bibinfo {author} {\bibfnamefont {J.~L.}\ \bibnamefont
  {MacCallum}}, \bibinfo {author} {\bibfnamefont {M.~J.}\ \bibnamefont
  {Hinner}}, \bibinfo {author} {\bibfnamefont {S.~J.}\ \bibnamefont {Marrink}},
  \ and\ \bibinfo {author} {\bibfnamefont {D.~P.}\ \bibnamefont {Tieleman}},\
  }\href@noop {} {\bibfield  {journal} {\bibinfo  {journal} {J. Am. Chem.
  Soc.}\ }\textbf {\bibinfo {volume} {131}},\ \bibinfo {pages} {12714}
  (\bibinfo {year} {2009})}\BibitemShut {NoStop}%
\bibitem [{\citenamefont {Bennett}\ and\ \citenamefont
  {Tieleman}(2012)}]{bennett.JLR.12}%
  \BibitemOpen
  \bibfield  {author} {\bibinfo {author} {\bibfnamefont {W.~F.~D.}\
  \bibnamefont {Bennett}}\ and\ \bibinfo {author} {\bibfnamefont {D.~P.}\
  \bibnamefont {Tieleman}},\ }\href@noop {} {\bibfield  {journal} {\bibinfo
  {journal} {J. Lipid Res.}\ }\textbf {\bibinfo {volume} {53}},\ \bibinfo
  {pages} {421} (\bibinfo {year} {2012})}\BibitemShut {NoStop}%
\bibitem [{\citenamefont {Ben-Shaul}(1995)}]{BenShaul.book.95}%
  \BibitemOpen
  \bibfield  {author} {\bibinfo {author} {\bibfnamefont {A.}~\bibnamefont
  {Ben-Shaul}},\ }in\ \href@noop {} {\emph {\bibinfo {booktitle} {Handbook of
  biological physics}}},\ \bibinfo {editor} {edited by\ \bibinfo {editor}
  {\bibfnamefont {R.}~\bibnamefont {Lipowsky}}\ and\ \bibinfo {editor}
  {\bibfnamefont {E.}~\bibnamefont {Sackmann}}}\ (\bibinfo  {publisher}
  {Elsevier},\ \bibinfo {year} {1995})\ pp.\ \bibinfo {pages}
  {359--401}\BibitemShut {NoStop}%
\bibitem [{\citenamefont {Bychuk}\ and\ \citenamefont
  {O'Shaughnessy}(1995)}]{BychukPrl95}%
  \BibitemOpen
  \bibfield  {author} {\bibinfo {author} {\bibfnamefont {O.~V.}\ \bibnamefont
  {Bychuk}}\ and\ \bibinfo {author} {\bibfnamefont {B.}~\bibnamefont
  {O'Shaughnessy}},\ }\href@noop {} {\bibfield  {journal} {\bibinfo  {journal}
  {Phys. Rev. Lett.}\ }\textbf {\bibinfo {volume} {74}},\ \bibinfo {pages}
  {1795} (\bibinfo {year} {1995})}\BibitemShut {NoStop}%
\bibitem [{\citenamefont {Kitson}\ \emph {et~al.}(1994)\citenamefont {Kitson},
  \citenamefont {Thewalt}, \citenamefont {Lafleur},\ and\ \citenamefont
  {Bloom}}]{kitson.biochem.94}%
  \BibitemOpen
  \bibfield  {author} {\bibinfo {author} {\bibfnamefont {N.}~\bibnamefont
  {Kitson}}, \bibinfo {author} {\bibfnamefont {J.}~\bibnamefont {Thewalt}},
  \bibinfo {author} {\bibfnamefont {M.}~\bibnamefont {Lafleur}}, \ and\
  \bibinfo {author} {\bibfnamefont {M.}~\bibnamefont {Bloom}},\ }\href@noop {}
  {\bibfield  {journal} {\bibinfo  {journal} {Biochemistry}\ }\textbf {\bibinfo
  {volume} {33}},\ \bibinfo {pages} {6707} (\bibinfo {year}
  {1994})}\BibitemShut {NoStop}%
\bibitem [{\citenamefont {Cordier}\ \emph {et~al.}(2008)\citenamefont
  {Cordier}, \citenamefont {Tournilhac}, \citenamefont {Souli{\'e}-Ziakovic},\
  and\ \citenamefont {Leibler}}]{cordier2008self}%
  \BibitemOpen
  \bibfield  {author} {\bibinfo {author} {\bibfnamefont {P.}~\bibnamefont
  {Cordier}}, \bibinfo {author} {\bibfnamefont {F.}~\bibnamefont {Tournilhac}},
  \bibinfo {author} {\bibfnamefont {C.}~\bibnamefont {Souli{\'e}-Ziakovic}}, \
  and\ \bibinfo {author} {\bibfnamefont {L.}~\bibnamefont {Leibler}},\
  }\href@noop {} {\bibfield  {journal} {\bibinfo  {journal} {Nature}\ }\textbf
  {\bibinfo {volume} {451}},\ \bibinfo {pages} {977} (\bibinfo {year}
  {2008})}\BibitemShut {NoStop}%
\end{thebibliography}%




\begin{thebibliography}{6}%
\makeatletter
\providecommand \@ifxundefined [1]{%
 \@ifx{#1\undefined}
}%
\providecommand \@ifnum [1]{%
 \ifnum #1\expandafter \@firstoftwo
 \else \expandafter \@secondoftwo
 \fi
}%
\providecommand \@ifx [1]{%
 \ifx #1\expandafter \@firstoftwo
 \else \expandafter \@secondoftwo
 \fi
}%
\providecommand \natexlab [1]{#1}%
\providecommand \enquote  [1]{``#1''}%
\providecommand \bibnamefont  [1]{#1}%
\providecommand \bibfnamefont [1]{#1}%
\providecommand \citenamefont [1]{#1}%
\providecommand \href@noop [0]{\@secondoftwo}%
\providecommand \href [0]{\begingroup \@sanitize@url \@href}%
\providecommand \@href[1]{\@@startlink{#1}\@@href}%
\providecommand \@@href[1]{\endgroup#1\@@endlink}%
\providecommand \@sanitize@url [0]{\catcode `\\12\catcode `\$12\catcode
  `\&12\catcode `\#12\catcode `\^12\catcode `\_12\catcode `\%12\relax}%
\providecommand \@@startlink[1]{}%
\providecommand \@@endlink[0]{}%
\providecommand \url  [0]{\begingroup\@sanitize@url \@url }%
\providecommand \@url [1]{\endgroup\@href {#1}{\urlprefix }}%
\providecommand \urlprefix  [0]{URL }%
\providecommand \Eprint [0]{\href }%
\providecommand \doibase [0]{http://dx.doi.org/}%
\providecommand \selectlanguage [0]{\@gobble}%
\providecommand \bibinfo  [0]{\@secondoftwo}%
\providecommand \bibfield  [0]{\@secondoftwo}%
\providecommand \translation [1]{[#1]}%
\providecommand \BibitemOpen [0]{}%
\providecommand \bibitemStop [0]{}%
\providecommand \bibitemNoStop [0]{.\EOS\space}%
\providecommand \EOS [0]{\spacefactor3000\relax}%
\providecommand \BibitemShut  [1]{\csname bibitem#1\endcsname}%
\let\auto@bib@innerbib\@empty
\bibitem [{\citenamefont {Ferrario}\ \emph {et~al.}(1990)\citenamefont
  {Ferrario}, \citenamefont {Haughney}, \citenamefont {McDonald},\ and\
  \citenamefont {Klein}}]{ferrario.jcp.90}%
  \BibitemOpen
  \bibfield  {author} {\bibinfo {author} {\bibfnamefont {M.}~\bibnamefont
  {Ferrario}}, \bibinfo {author} {\bibfnamefont {M.}~\bibnamefont {Haughney}},
  \bibinfo {author} {\bibfnamefont {I.~R.}\ \bibnamefont {McDonald}}, \ and\
  \bibinfo {author} {\bibfnamefont {M.~L.}\ \bibnamefont {Klein}},\ }\href@noop
  {} {\bibfield  {journal} {\bibinfo  {journal} {J. Chem. Phys.}\ }\textbf
  {\bibinfo {volume} {93}},\ \bibinfo {pages} {5156} (\bibinfo {year}
  {1990})}\BibitemShut {NoStop}%
\bibitem [{\citenamefont {Luzar}\ and\ \citenamefont
  {Chandler}(1993)}]{luzar.jcp.93}%
  \BibitemOpen
  \bibfield  {author} {\bibinfo {author} {\bibfnamefont {A.}~\bibnamefont
  {Luzar}}\ and\ \bibinfo {author} {\bibfnamefont {D.}~\bibnamefont
  {Chandler}},\ }\href@noop {} {\bibfield  {journal} {\bibinfo  {journal} {J.
  Chem. Phys.}\ }\textbf {\bibinfo {volume} {98}},\ \bibinfo {pages} {8160}
  (\bibinfo {year} {1993})}\BibitemShut {NoStop}%
\bibitem [{\citenamefont {Luzar}\ and\ \citenamefont
  {Chandler}(1996)}]{luzar.prl.96}%
  \BibitemOpen
  \bibfield  {author} {\bibinfo {author} {\bibfnamefont {A.}~\bibnamefont
  {Luzar}}\ and\ \bibinfo {author} {\bibfnamefont {D.}~\bibnamefont
  {Chandler}},\ }\href@noop {} {\bibfield  {journal} {\bibinfo  {journal}
  {Phys. Rev. Lett.}\ }\textbf {\bibinfo {volume} {76}},\ \bibinfo {pages}
  {928} (\bibinfo {year} {1996})}\BibitemShut {NoStop}%
\bibitem [{\citenamefont {Torshin}\ \emph {et~al.}(2002)\citenamefont
  {Torshin}, \citenamefont {Weber},\ and\ \citenamefont
  {Harrison}}]{torshin.PE.02}%
  \BibitemOpen
  \bibfield  {author} {\bibinfo {author} {\bibfnamefont {I.~Y.}\ \bibnamefont
  {Torshin}}, \bibinfo {author} {\bibfnamefont {I.~T.}\ \bibnamefont {Weber}},
  \ and\ \bibinfo {author} {\bibfnamefont {R.~W.}\ \bibnamefont {Harrison}},\
  }\href@noop {} {\bibfield  {journal} {\bibinfo  {journal} {Protein
  engineering}\ }\textbf {\bibinfo {volume} {15}},\ \bibinfo {pages} {359}
  (\bibinfo {year} {2002})}\BibitemShut {NoStop}%
\bibitem [{\citenamefont {Vermeer}\ \emph {et~al.}(2007)\citenamefont
  {Vermeer}, \citenamefont {de~Groot}, \citenamefont {R\'eat}, \citenamefont
  {Milon},\ and\ \citenamefont {Czaplicki}}]{vermeer.EBJ.07}%
  \BibitemOpen
  \bibfield  {author} {\bibinfo {author} {\bibfnamefont {L.~S.}\ \bibnamefont
  {Vermeer}}, \bibinfo {author} {\bibfnamefont {B.~L.}\ \bibnamefont
  {de~Groot}}, \bibinfo {author} {\bibfnamefont {V.}~\bibnamefont {R\'eat}},
  \bibinfo {author} {\bibfnamefont {A.}~\bibnamefont {Milon}}, \ and\ \bibinfo
  {author} {\bibfnamefont {J.}~\bibnamefont {Czaplicki}},\ }\href@noop {}
  {\bibfield  {journal} {\bibinfo  {journal} {Eur. Biophys. J.}\ }\textbf
  {\bibinfo {volume} {36}},\ \bibinfo {pages} {919} (\bibinfo {year}
  {2007})}\BibitemShut {NoStop}%
\bibitem [{\citenamefont {Den~Otter}\ and\ \citenamefont
  {Shkulipa}(2007)}]{denOtter.BPJ.07}%
  \BibitemOpen
  \bibfield  {author} {\bibinfo {author} {\bibfnamefont {W.}~\bibnamefont
  {Den~Otter}}\ and\ \bibinfo {author} {\bibfnamefont {S.}~\bibnamefont
  {Shkulipa}},\ }\href@noop {} {\bibfield  {journal} {\bibinfo  {journal}
  {Biophys J.}\ }\textbf {\bibinfo {volume} {93}},\ \bibinfo {pages} {423}
  (\bibinfo {year} {2007})}\BibitemShut {NoStop}%
\end{thebibliography}

%
\end{document}